\documentclass[twocolumn]{aastex63}
\submitjournal{ApJ}
\begin{document}


\shorttitle{
Extended Ly$\alpha$ Emission Around $z\sim2-7$ LAEs
}

\shortauthors{
Kikuchihara et al.
}

\title{
SILVERRUSH. XI. Intensity Mapping for Ly$\alpha$ Emission Extending over\\
100-1000 comoving kpc around $z\sim2-7$ LAEs with Subaru HSC-SSP and CHORUS Data
}

\author[0000-0003-2449-6314]{Shotaro Kikuchihara}
\affiliation{Institute for Cosmic Ray Research, The University of Tokyo, 5-1-5 Kashiwanoha, Kashiwa, Chiba 277-8582, Japan}
\affiliation{Department of Astronomy, Graduate School of Science, The University of Tokyo, 7-3-1 Hongo, Bunkyo-ku, Tokyo 113-0033, Japan}
\email{skiku@icrr.u-tokyo.ac.jp}

\author[0000-0002-6047-430X]{Yuichi Harikane}
\affiliation{Institute for Cosmic Ray Research, The University of Tokyo, 5-1-5 Kashiwanoha, Kashiwa, Chiba 277-8582, Japan}
\affiliation{Department of Physics and Astronomy, University College London, Gower Street, London WC1E 6BT, UK}

\author[0000-0002-1049-6658]{Masami Ouchi}
\affiliation{Institute for Cosmic Ray Research, The University of Tokyo, 5-1-5 Kashiwanoha, Kashiwa, Chiba 277-8582, Japan}
\affiliation{National Astronomical Observatory of Japan, 2-21-1 Osawa, Mitaka, Tokyo 181-8588, Japan}
\affiliation{Graduate University for Advanced Studies (SOKENDAI), 2-21-1 Osawa, Mitaka, Tokyo 181-8588, Japan}
\affiliation{Kavli Institute for the Physics and Mathematics of the Universe (Kavli IPMU, WPI), The University of Tokyo, 5-1-5 Kashiwanoha, Kashiwa, Chiba 277-8583, Japan}

\author[0000-0001-9011-7605]{Yoshiaki Ono}
\affiliation{Institute for Cosmic Ray Research, The University of Tokyo, 5-1-5 Kashiwanoha, Kashiwa, Chiba 277-8582, Japan}

\author{Takatoshi Shibuya}
\affiliation{Kitami Institute of Technology, 165 Koen-cho, Kitami, Hokkaido 090-8507, Japan}

\author{Ryohei Itoh}
\affiliation{Institute for Cosmic Ray Research, The University of Tokyo, 5-1-5 Kashiwanoha, Kashiwa, Chiba 277-8582, Japan}
\affiliation{Department of Physics, Graduate School of Science, The University of Tokyo, 7-3-1 Hongo, Bunkyo-ku, Tokyo 113-0033, Japan}

\author{Ryota Kakuma}
\affiliation{Institute for Cosmic Ray Research, The University of Tokyo, 5-1-5 Kashiwanoha, Kashiwa, Chiba 277-8582, Japan}
\affiliation{Department of Astronomy, Graduate School of Science, The University of Tokyo, 7-3-1 Hongo, Bunkyo-ku, Tokyo 113-0033, Japan}

\author[0000-0002-7779-8677]{Akio K. Inoue}
\affiliation{Waseda Research Institute for Science and Engineering, Faculty of Science and Engineering, Waseda University, 3-4-1 Okubo, Shinjuku, Tokyo 169-8555, Japan}
\affiliation{Department of Physics, School of Advanced Science and Engineering, Faculty of Science and Engineering, Waseda University, 3-4-1 Okubo, Shinjuku, Tokyo 169-8555, Japan}

\author[0000-0002-3801-434X]{Haruka Kusakabe}
\affiliation{Observatoire de Gen\`{e}ve, Universit\'{e} de Gen\`{e}ve, 51 chemin de P\'{e}gase, 1290 Versoix, Switzerland}

\author[0000-0002-2597-2231]{Kazuhiro Shimasaku}
\affiliation{Department of Astronomy, Graduate School of Science, The University of Tokyo, 7-3-1 Hongo, Bunkyo-ku, Tokyo 113-0033, Japan}
\affiliation{Research Center for the Early Universe, Graduate School of Science, The University of Tokyo, 7-3-1 Hongo, Bunkyo-ku, Tokyo 113-0033, Japan}

\author[0000-0002-8857-2905]{Rieko Momose}
\affiliation{Department of Astronomy, Graduate School of Science, The University of Tokyo, 7-3-1 Hongo, Bunkyo-ku, Tokyo 113-0033, Japan}

\author[0000-0001-6958-7856]{Yuma Sugahara}
\affiliation{National Astronomical Observatory of Japan, 2-21-1 Osawa, Mitaka, Tokyo 181-8588, Japan}
\affiliation{Waseda Research Institute for Science and Engineering, Faculty of Science and Engineering, Waseda University, 3-4-1 Okubo, Shinjuku, Tokyo 169-8555, Japan}

\author[0000-0003-3214-9128]{Satoshi Kikuta}
\affiliation{Center for Computational Sciences, University of Tsukuba, Ten-nodai, 1-1-1 Tsukuba, Ibaraki 305-8577, Japan}

\author[0000-0002-6186-5476]{Shun Saito}
\affiliation{Institute for Multi-messenger Astrophysics and Cosmology, Department of Physics, Missouri University of Science and Technology, 1315 N. Pine Street, Rolla, MO 65409, USA}

\author[0000-0001-5493-6259]{Nobunari Kashikawa}
\affiliation{Department of Astronomy, Graduate School of Science, The University of Tokyo, 7-3-1 Hongo, Bunkyo-ku, Tokyo 113-0033, Japan}
\affiliation{Research Center for the Early Universe, Graduate School of Science, The University of Tokyo, 7-3-1 Hongo, Bunkyo-ku, Tokyo 113-0033, Japan}

\author[0000-0003-2273-9415]{Haibin Zhang}
\affiliation{Department of Astronomy, Tsinghua University, No. 1 Qinghuayuan, Beijing 100084, China}

\author[0000-0003-1700-5740]{Chien-Hsiu Lee}
\affiliation{NSF’s National Optical-Infrared Astronomy Research Laboratory, Tucson, AZ 85719, USA}


\begin{abstract}
We conduct intensity mapping to probe for extended diffuse Ly$\alpha$ emission around Ly$\alpha$ emitters (LAEs) at $z\sim2-7$, exploiting very deep ($\sim26$ mag at $5\sigma$) and large-area ($\sim4.5$ deg$^2$) Subaru/Hyper Suprime-Cam narrow-band (NB) images and large LAE catalogs consisting of a total of 1781 LAEs at $z=2.2$, $3.3$, $5.7$, and $6.6$ obtained by the HSC-SSP SILVERRUSH and CHORUS projects. We calculate the spatial correlations of these LAEs with $\sim1-2$ billion pixel flux values of the NB images, deriving the average Ly$\alpha$ surface brightness (${\rm SB_{Ly\alpha}}$) radial profiles around the LAEs. By carefully estimating systematics such as fluctuations of sky background and point spread functions, we detect diffuse Ly$\alpha$ emission ($\sim10^{-20}-10^{-19}$ erg s$^{-1}$ cm$^{-2}$ arcsec$^{-2}$) at $100-1000$ comoving kpc around $z=3.3$ LAEs at the $4.1\sigma$ level and tentatively ($\sim2\sigma$) at the other redshifts, beyond the virial radius of a dark-matter halo with a mass of $10^{11}\ M_\odot$. While the observed ${\rm SB_{Ly\alpha}}$ profiles have similar amplitudes at $z=2.2-6.6$ within the uncertainties, the intrinsic ${\rm SB_{Ly\alpha}}$ profiles (corrected for the cosmological dimming effect) increase toward high redshifts. This trend may be explained by increasing hydrogen gas density due to the evolution of the cosmic volume. Comparisons with theoretical models suggest that extended Ly$\alpha$ emission around a LAE is powered by resonantly scattered Ly$\alpha$ photons in the CGM and IGM that originates from the inner part of the LAE, and/or neighboring galaxies around the LAE.
\end{abstract}



\section{Introduction} \label{sec:intro}

The gas surrounding a galaxy, called the circumgalactic medium (CGM), falls into the galaxy and triggers star formation activity, and is subsequently ejected from the galaxy due to outflows \citep[e.g.,][]{Tumlinson+17,Peroux-Howk20}.
The hydrogen gas inside the CGM can be traced by Lyman-alpha (Ly$\alpha$) emission, which is observed as a Ly$\alpha$ halo (LAH).
Therefore, observing LAHs is key to understanding the properties and kinematics of the CGM, and eventually provide information on galaxy formation and evolution.

Many studies have detected LAHs around nearby galaxies \citep[e.g.,][]{Ostlin+09,Hayes+13,Hayes+14}.
At high-redshift ($z>2$), meanwhile, LAHs have been identified mainly around massive galaxies, such as Lyman break galaxies \citep[e.g.,][]{Hayashino+04,Swinbank+07,Steidel+11} and quasars \citep[e.g.,][]{Goto+09,Cantalupo+14,Martin+14,Borisova+16,Arrigoni-Battaia+19,Kikuta+19,Zhang+20}.
However, it remains difficult to detect diffuse emission around less massive star-forming galaxies (SFGs), such as Ly$\alpha$ emitters (LAEs), at high redshift, due to their faintness and sensitivity limits.

To overcome this difficulty, \citet{Rauch+08}, for example, performed a very deep (92 hr) long-slit observation with the ESO Very Large Telescope (VLT)/FOcal Reducer and low dispersion Spectrograph 2 (FORS2) that reached a $1\sigma$ surface brightness (SB) detection limit of $8\times10^{-20}$ erg cm$^{-2}$ s$^{-1}$ arcsec$^{-2}$.
They investigated 27 LAEs at $z=2.67-3.75$, identifying Ly$\alpha$ emission extending over 26 physical kpc (pkpc) around one of the LAEs.
Individual detections of many high redshift LAHs have been enabled by the advent of the Multi-Unit Spectroscopic Explorer (MUSE) installed on the VLT.
Recently, \citet{Leclercq+17} identified individual LAHs around 145 LAEs at $z=3-6$ in the Hubble Ultra Deep Field (HUDF) with the VLT/MUSE \citep[see also][]{Wisotzki+16}.
Their data reached a SB limit of $\lesssim10^{-19}$ erg s$^{-1}$ cm$^{-2}$ arcsec$^{-2}$ at radii of $>10$ pkpc.
The magnification effect caused by gravitational lensing was also utilized in conjunction with the MUSE for studying individual LAHs \citep{Patricio+16,Smit+17,Claeyssens+19}.

A stacking method has been widely used to obtain averaged LAH profiles with high signal-to-noise (S/N) ratios \citep[e.g.,][]{Matsuda+12,Momose+14,Momose+16,Xue+17,Wisotzki+18,Wu+20}.
For example, \citet{Momose+14} stacked $>100$ narrow-band (NB) images around LAEs at $z=2.2-6.6$ using Subaru Telescope/Suprime-Cam (SC) data, investigating LAH profiles up to $\sim50$ pkpc radial scales.
\citet{Matsuda+12} and \citet{Momose+16} investigated the LAH size dependence on LAE properties, such as Ly$\alpha$ luminosity, rest-frame ultra-violet (UV) magnitude, and overdensity, at $z=3.1$ and 2.2, respectively.
We note that some studies \citep[e.g.][]{Bond+10,Feldmeier+13,Jiang+13} reported no evidence of extended Ly$\alpha$ emission at $z>2$.

Another approach is the intensity mapping technique (\citealt{Kovetz+17} for a review; see also \citealt{Carilli11,Gong+11,Silva+13,Pullen+14,Comaschi-Ferrara16a,Comaschi-Ferrara+16b,Li+16,Fonseca+17}), which utilizes cross-correlation functions between objects and their emission or absorption spectra.
This technique enables us to detect signals from targeted galaxies with a high S/N ratio by efficiently estimating and removing contaminating signals from foreground interlopers.
\citet{Croft+16,Croft+18} derived cross-correlation functions between the Ly$\alpha$ emission and quasar positions at $z=2-3.5$ using data from the Sloan Digital Sky Survey \citep[SDSS;][]{Eisenstein+11} Baryon Oscillation Spectroscopic Survey \citep[BOSS;][]{Dawson+13}. This work enabled the detection of positive signals up to a $\sim15\ h^{-1}$ comoving Mpc (cMpc) radial scale.

Recently, \citet{Kakuma+21} applied the intensity mapping technique to LAEs at $z=5.7$ and 6.6 using Subaru/Hyper Suprime-Cam (HSC) data.
They tentatively identified very diffuse ($\sim10^{-20}\ {\rm erg\ s^{-1}\ cm^{-2}\ arcsec^{-2}}$) Ly$\alpha$ emission extended around the LAEs over the radial scale of the virial radius ($R_{\rm vir}$) of a dark-matter halo (DMH).
Although their finding has shed light on the potential existence of Ly$\alpha$-emitting hydrogen gas beyond $R_{\rm vir}$ around LAEs in the reionization epoch, it remains an open question whether LAEs host such extended structures even at lower redshifts ($z<5$).
In this paper, we exploit NB images offered in two Subaru/HSC surveys, the Cosmic HydrOgen Reionization Unveiled with Subaru \citep[CHORUS;][]{Inoue+20} and the Subaru Strategic Program \citep[HSC-SSP;][]{Aihara+19},
which enable us to trace Ly$\alpha$ emission from LAEs across $z=2.2-6.6$.
Our goal is to systematically investigate diffuse Ly$\alpha$ emission extended beyond $R_{\rm vir}$ around $z=2.2-6.6$ LAEs, taking advantage of the intensity mapping technique and ultra-deep images from the CHORUS and HSC-SSP projects.

Another open question about extended Ly$\alpha$ emission is its physical origin (see review by \citealt{Ouchi+20} and Figure 15 of \citealt{Momose+16}).
Theoretical studies have suggested several physical processes producing a LAH around a galaxy, which can be attributed mainly to 1) resonant scattering and 2) {\it in-situ} production. 1) Resonant scattering: Ly$\alpha$ photons are produced in the interstellar medium (ISM) of the galaxy, and then resonantly scattered by neutral hydrogen gas while escaping the galaxy into the CGM and intergalactic medium (IGM) \citep[e.g.,][]{Laursen-Sommer-Larsen07,Laursen+11,Steidel+11,Zheng+11,Dijkstra-Kramer12,Jeeson-Daniel+12,Verhamme+12,Kakiichi-Dijkstra18,Smith+18,Smith+19,Garel+21}.
2) {\it In-situ} production: Ly$\alpha$ photons are produced not inside the galaxy but in the CGM. This can be further classified into three processes: i) recombination, ii) collisional excitation, and iii) satellite galaxies. i) Recombination: ionizing radiation from the galaxy or extragalactic background (UVB) photoionizes the hydrogen gas in the CGM, which in turn emit Ly$\alpha$ emission via recombination \citep[`fluorescence'; e.g.,][]{Furlanetto+05,Cantalupo+05,Kollmeier+10,Lake+15,Mas-Ribas-Dijkstra16,Gallego+18,Mas-Ribas+17c}.
ii) Collisional excitation: Hydrogen gas in the CGM is compressively heated by shocks and then emit Ly$\alpha$ photons by converting its gravitational energy into Ly$\alpha$ emission when they accrete onto the galaxy \citep[`gravitational cooling' or `cold stream'; e.g.,][]{Haiman+00,Fardal+01,Goerdt+10,Faucher-Gigure+10,Rosdahl-Blaizot12,Lake+15}.
iii) Satellite galaxies: Ly$\alpha$ emission is produced by star formation (SF) in unresolved dwarfs surrounding the galaxy \citep[`satellite galaxies'; e.g.,][]{Mas-Ribas+17a,Mas-Ribas+17c}.
To recognize which process plays a major role, \citet{Kakuma+21} compared the observed Ly$\alpha$ SB profiles against the prediction by \citet{Zheng+11} that considers resonant scattering, although large uncertainties of the data prevented to draw a conclusion.
Since their comparison was limited to the case for resonant scattering at $z=5.7$, we need to systematically investigate various origins by utilizing multiple models at multiple redshifts to pin down the key origin(s), in the similar way as \citet{Byrohl+20} and \citet{Mitchell+21}.

This paper is organized as follows.
Our data are described in Section \ref{sec:data}.
In Section \ref{sec:im}, we use the intensity mapping technique to derive the cross-correlation SB of Ly$\alpha$ emission around the LAEs.
We discuss the redshift evolution and physical origins of extended Ly$\alpha$ emission in Section \ref{sec:discussion} and summarize our findings in Section \ref{sec:summary}.

Throughout this paper, magnitudes are given in the AB system \citep{Oke-Gunn83}.
We adopt the concordance cosmology with $\Omega_{\rm m,0}=0.7$, $\Omega_{\Lambda,0}=0.3$, and $H_0=70\ {\rm km\ s^{-1}\ Mpc^{-1}}$, where $1\arcsec$ corresponds to transverse sizes of (8.3, 7.5, 5.9, 5.4) pkpc and (26, 32, 39, 41) comoving kpc (ckpc) at $z=(2.2,\ 3.3,\ 5.7,\ 6.6)$.

\section{Data} \label{sec:data}

In this Section, we describe the images and sample catalogs used for our analyses.
All the images and catalogs are based on Subaru/HSC data.

\subsection{Images} \label{subsec:images}

We use NB and broad-band (BB) imaging data that were obtained in two Subaru/HSC surveys, the HSC-SSP and the CHORUS.
The HSC-SSP and CHORUS data were obtained in March 2014-January 2018 and January 2017-December 2018, respectively.
We specifically use the internal data of the S18A release.
The HSC-SSP survey is a combination of three layers: Wide, Deep, and UltraDeep (UD).
We use the UD layer images in the fields of the Cosmological Evolution Survey \citep[UD-COSMOS;][]{Scoville+07} and Subaru/XMM Deep Survey \citep[UD-SXDS;][]{Sekiguchi+05}, because the wide survey areas ($\sim2$ deg$^2$ for each field) and deep imaging (the $5\sigma$ limiting magnitudes are $\sim26$ mag in a $2\arcsec$-diameter aperture) in these fields are advantageous for the detection of very diffuse Ly$\alpha$ emission.
The CHORUS data were obtained over the UD-COSMOS field.
The HSC-SSP and CHORUS data were reduced with the HSC pipeline v6.7 \citep{Bosch+18}.

\begin{figure}
\epsscale{1.1}
\plotone{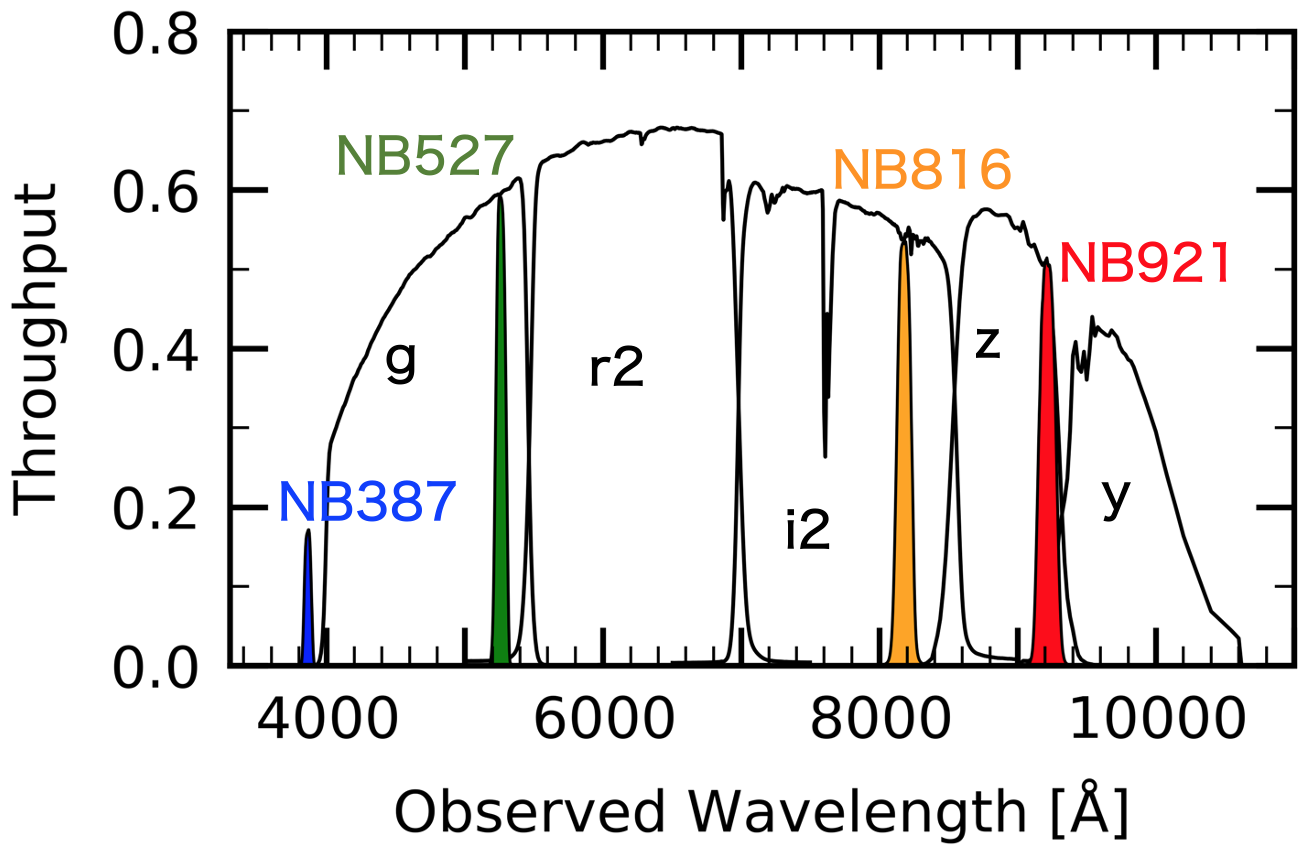}
\caption{Effective response curves of the HSC NB filters: NB387 (blue), NB527 (green), NB816 (orange), and NB921 (red), and BB filters: $g$, $r2$, $i2$, $z$, and $y$ (black). These response curves include the quantum efficiency of the HSC CCD, transmittance of the dewar window and of the Primary Focus Unit, reflectivity of the Primary Mirror, and airmass at the Telescope site.}
\label{fig:filters}
\end{figure}

The HSC-SSP program of S18A provides the data of two NB (NB816 and NB921) filters in the UD-COSMOS and UD-SXDS fields, while the CHORUS images are offered in four NB (NB387, NB527, NB718, and NB973) filters in the UD-COSMOS field.
In this work, we present the results in the NB387, NB527, NB816, and NB921 filters.
The NB718 and NB973 filters are not used in the following sections, because the number of LAEs and the image depths are not sufficient to detect diffuse Ly$\alpha$ emission.
The NB387, NB527, NB816, and NB921 filters are centered at 3863, 5260, 8177, and 9215 {\AA} with the full widths at half maximum (FWHMs) of 55, 79, 113, and 135 {\AA}, respectively, which cover the observed wavelengths of Ly$\alpha$ emission from $z=2.178\pm0.023$, $3.327\pm0.032$, $5.726\pm0.046$, and $6.580\pm0.056$, respectively.
Five BB ($g$-, $r2$-, $i2$-, $z$-, and $y$-band) filters are also available in both HSC-SSP and CHORUS.
Figure \ref{fig:filters} shows the NB and BB filter throughputs, and Table \ref{tab:filters_and_image_summary} summarizes the images and filters.

\begin{deluxetable*}{ccccccccc}
\tablecaption{Summary of the NB filters and images.
\label{tab:filters_and_image_summary}}
\tablehead{
\colhead{} & \colhead{} & \colhead{} & \colhead{} & \twocolhead{UD-COSMOS} & & \twocolhead{UD-SXDS} \\ \cline{5-6} \cline{8-9}
\colhead{NB} & \colhead{$\lambda_{\rm c}$ (\AA)} & \colhead{FWHM (\AA)} & \colhead{$z_{\rm LAE}$} & \colhead{Area (deg$^2$)} & \colhead{$m_{{\rm NB},5\sigma}$ (mag)} & & \colhead{Area (deg$^2$)} & \colhead{$m_{{\rm NB},5\sigma}$ (mag)} \\
\colhead{(1)} & \colhead{(2)} & \colhead{(3)} & \colhead{(4)} & \colhead{(5)} & \colhead{(6)} & & \colhead{(7)} & \colhead{(8)}
}
\startdata
NB387 (CHORUS) & 3863 &  55 & $2.178\pm0.023$ & 1.561 & 25.67 & & --- & --- \\
NB527 (CHORUS) & 5260 &  79 & $3.327\pm0.032$ & 1.613 & 26.39 & & --- & --- \\
NB816 (HSC-SSP) & 8177 & 113 & $5.726\pm0.046$ & 2.261 & 25.75 & & 2.278 & 25.61 \\
NB921 (HSC-SSP) & 9215 & 135 & $6.580\pm0.056$ & 2.278 & 25.48 & & 2.278 & 25.31
\enddata
\tablecomments{Columns: (1) NB filter. (2)-(3) Central wavelength ($\lambda_{\rm c}$) and FWHM of the NB filter transmission curve. (4) Redshift range of the LAEs whose Ly$\alpha$ emission enters the NB filter. (5)-(6) Effective area and $5\sigma$ limiting NB magnitude ($m_{{\rm NB},5\sigma}$) in the UD-COSMOS field. The $m_{{\rm NB},5\sigma}$ value is measured with a $2\arcsec$-diameter aperture in each patch, and then averaged over the field. See \citet{Inoue+20} and \citet{Hayashi+20} for the spatial variance of $m_{{\rm NB},5\sigma}$. The $m_{{\rm NB},5\sigma}$ value in NB387 is corrected for the systematic zero-point offset by 0.45 mag (see Section \ref{subsec:lae}). (7)-(8) Same as Columns (5)-(6), but for the UD-SXDS field. The values in this Table are cited from \citet{Ono+21}.}
\end{deluxetable*}

Bright sources in the NB and BB images must be masked since they contaminate diffuse emission.
We thus mask pixels flagged with either {\tt DETECT} or {\tt BRIGHT\_OBJECT} using the masks provided by the HSC pipeline (termed {\it original} masks).
A pixel is flagged with {\tt DETECT} or {\tt BRIGHT\_OBJECT} when the pixel is covered by a detected ($\geq5\sigma$) object or is affected by nearby bright sources, respectively.
However, because a part of bright sources were missed in the {\it original} masks due to bad photometry, the HSC-SSP team offered new masks that mitigated this problem (hereafter termed {\it revised} masks).\footnote{\url{https://hsc-release.mtk.nao.ac.jp/doc/index.php/bright-star-masks-2/}}
We adopt the {\it revised} masks in addition to the {\it original} masks to flag {\tt BRIGHT\_OBJECT}.
We use the {\it revised} $g$-, $r2$-, $z$-, and $y$-band masks for NB387, NB527, NB816, and NB921 images, respectively, because the {\it revised} masks are offered only in the BB filters.
For the BB images, we use the {\it revised} masks defined for each BB filter.
We visually confirm that these criteria successfully cover bright sources and contaminants in the images.

\subsection{LAE Samples} \label{subsec:lae}

We use the LAE catalog constructed by \citet{Ono+21} as a part of the Systematic Identification of LAEs for Visible Exploration and Reionization Research Using Subaru HSC (SILVERRUSH) project (\citealt{Ouchi+18}, see also \citealt{Shibuya+18a,Shibuya+18b,Konno+18,Harikane+18b,Inoue+18,Higuchi+18,Harikane+19,Kakuma+21}).
\citet{Ono+21} selected LAE candidates based on color and removed contaminants by a convolutional neural network (CNN) and visual inspection.
Their final catalog includes (542, 959, 395, 150) LAEs at $z=(2.2$, 3.3, 5.7, 6.6) in the UD-COSMOS field, and (560, 75) LAEs at $z=(5.7$, 6.6) in the UD-SXDS field.

\begin{figure*}
\epsscale{1.1}
\plotone{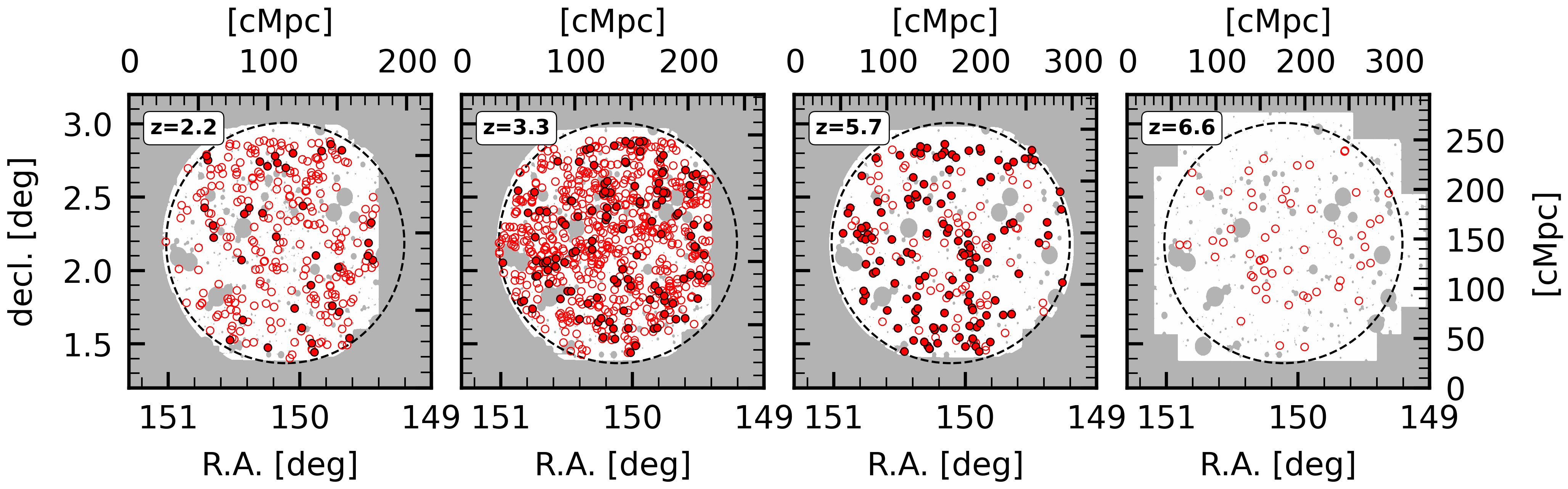}
\caption{Sky distributions of the LAEs in the UD-COSMOS field at $z=2.2$, 3.3, 5.7, and 6.6, from left to right. The open red circles show the positions of the LAEs included in the {\it all} sample but not in the {\it bright} subsample, while the filled red circles indicate those of the {\it bright} subsample LAEs (i.e., {\it all} sample LAEs are represented by the open$+$filled red circles). We use the LAEs inside the black dashed circles. The background white shades show the NB387, NB527, NB816, and NB921 images, from left to right. The gray shades show the regions where the pixel is masked or an UD image is not offered. Note that the distribution of the {\it bright} subsample at $z=6.6$ is not displayed, because we treat the {\it all} sample also as the {\it bright} subsample at $z=6.6$.}
\label{fig:skypositions_udcosmos}
\end{figure*}

\begin{figure}
\epsscale{1.1}
\plotone{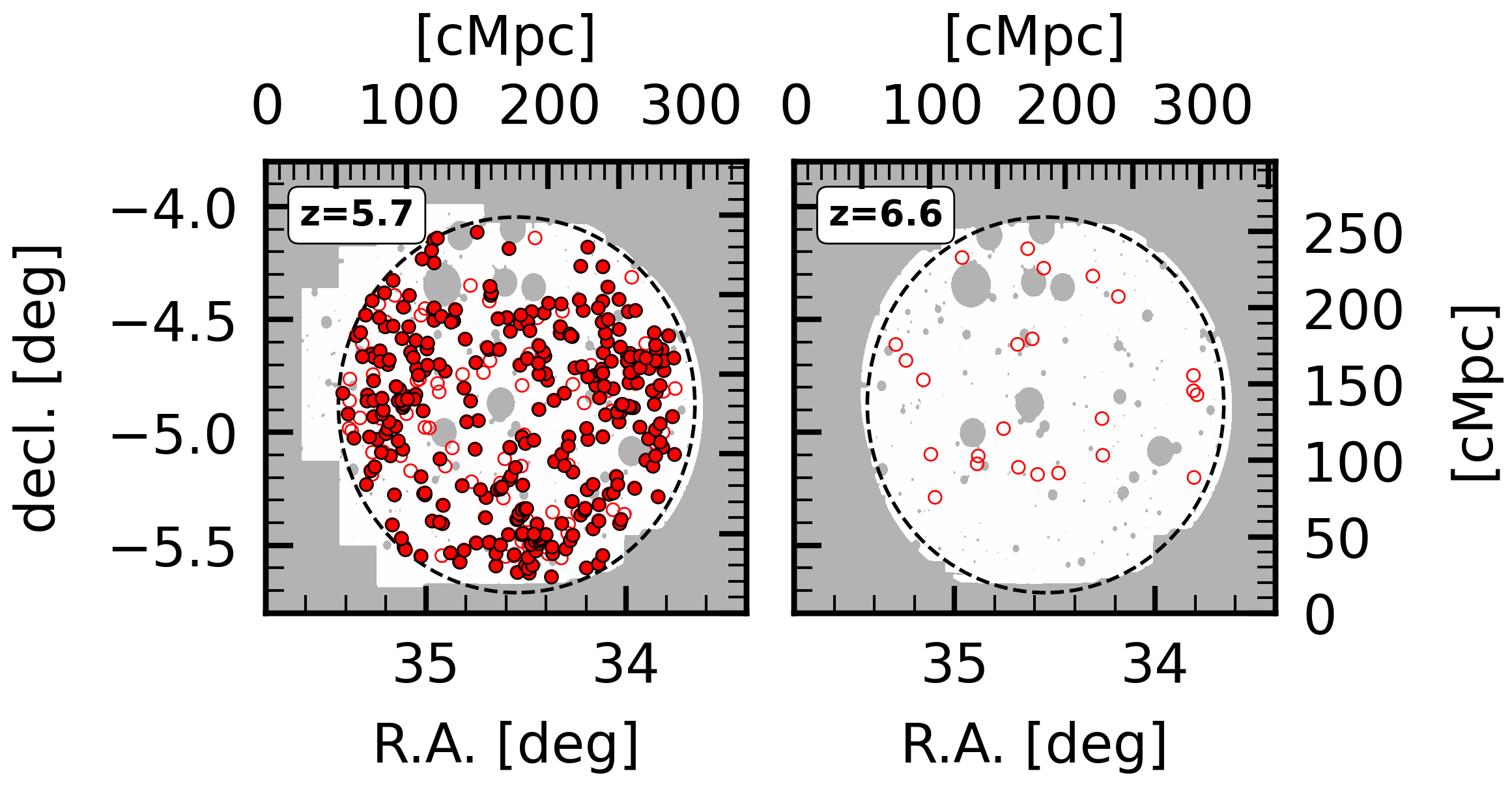}
\caption{Same as Figure \ref{fig:skypositions_udcosmos}, but for the UD-SXDS field at $z=5.7$ (left) and 6.6 (right).}
\label{fig:skypositions_udsxds}
\end{figure}

The NB images of the UD-COSMOS and UD-SXDS fields are deepest at the center and become shallower toward the edges \citep{Hayashi+20,Inoue+20}.
We thus exclude LAEs outside of the boundaries that are shown with the black dashed circles in Figures \ref{fig:skypositions_udcosmos} and \ref{fig:skypositions_udsxds}.

We estimate the Ly$\alpha$ line luminosities ($L_{\rm Ly\alpha}$) following \citet[see also \citealt{Itoh+18}]{Shibuya+18a}.
First, we measure the NB (BB) magnitudes $m_{\rm NB}$ ($m_{\rm BB}$) of the LAEs at $z=2.2$, 3.3, 5.7, an 6.6 in the NB387 ($g$-band), NB527 ($r2$-band), NB816 ($z$-band), and NB921 ($y$-band) filters, respectively.
The magnitudes are measured with a $2\arcsec$-diameter aperture because it efficiently covers the point spread function (PSF), whose FWHM is $0\farcs8-1\farcs1$ \citep{Ono+21}.
The NB387 magnitudes are corrected for the systematic zero-point offset by 0.45 mag, following the recommendation by the HSC-SSP team.\footnote{\url{https://hsc-release.mtk.nao.ac.jp/doc/index.php/known-problems-2/\#hsc-link-10}}
Next, we follow \citet{Shibuya+18a} in our derivation of the Ly$\alpha$ line fluxes ($f_{\rm Ly\alpha}$) from $m_{\rm NB}$ and $m_{\rm BB}$, assuming a flat UV continuum and the IGM attenuation model taken from \citet{Inoue+14}.
Lastly, the values of $L_{\rm Ly\alpha}$ are derived via $L_{\rm Ly\alpha}=4\pi d_{\rm L}(z_{\rm LAE})^2f_{\rm Ly\alpha}$, where $d_{\rm L}(z_{\rm LAE})$ denotes the luminosity distance to the LAE at redshift $z_{\rm LAE}$.

Although the completeness of the LAEs is as high as $\gtrsim90$ \% at $m_{\rm NB}\lesssim24.5$ in the CNN of \citet{Ono+21}, faint LAEs may be missed in the observations and selection.
To ensure completeness, we use only LAEs whose $L_{\rm Ly\alpha}$ values are larger than the modes (peaks) of the $L_{\rm Ly\alpha}$ histograms, which are represented as $L_{\rm Ly\alpha}^{\rm min}$ in Table \ref{tab:samples}.
This sample, termed the {\it all} sample, consists of (289, 762, 210, 56) LAEs at $z=(2.2$, 3.3, 5.7, 6.6) and (393, 24) LAEs at $z=(5.7$, 6.6) in the UD-COSMOS and UD-SXDS fields, respectively.
To accurately compare the LAEs of different redshifts at similar $L_{\rm Ly\alpha}$ values, we further exclude faint LAEs from the {\it all} sample such that the mean $L_{\rm Ly\alpha}$ values are equal to $10^{42.9}$ erg s$^{-1}$ at each redshift.
This selection results in (37, 123, 125) LAEs at $z=(2.2$, 3.3, 5.7) and 313 LAEs at $z=5.7$ in the UD-COSMOS and UD-SXDS fields, respectively, which we hereafter refer to as the {\it bright} subsample.
At $z=6.6$, since the mean $L_{\rm Ly\alpha}$ values of the {\it all} sample are $10^{43.0}$ erg s$^{-1}$ in both the UD-COSMOS and UD-SXDS fields, we also use the {\it all} sample as the {\it bright} subsample.
In summary, we use a total of 1781 and 717 LAEs in the UD-COSMOS$+$UD-SXDS fields as the {\it all} sample and the {\it bright} subsample, respectively.
The $L_{\rm Ly\alpha}$ values and sample sizes are summarized in Table \ref{tab:samples}.
The sky distributions of the LAEs in the UD-COSMOS and UD-SXDS fields are presented in Figures \ref{fig:skypositions_udcosmos} and \ref{fig:skypositions_udsxds}, respectively.

A major update of our data compared to \citet{Kakuma+21} is that we add new 1051 LAEs at $z=2.2$ and 3.3 using the CHORUS data.
The catalog of $z=5.7$ and 6.6 LAEs are also updated in that we use the latest catalog constructed by \citet{Ono+21} based on HSC-SSP S18A images, while \citet{Kakuma+21} used the S16A catalog taken from \citet{Shibuya+18a}.
Although the number of the LAEs increased owing to the improved limiting magnitudes, we excluded faint LAEs from those included in \citet{Ono+21}, which resulting in comparable numbers of the LAEs between \citet{Kakuma+21}'s and our {\it all} samples.
The sky and $L_{\rm Ly\alpha}$ distributions are also similar between these two samples.
We use the same images as \citet[i.e., those taken from the HSC-SSP S18A release]{Kakuma+21} in NB816 and NB921, while our masking prescription may be slightly different.

\begin{deluxetable*}{cccccccccccccccc}
\tablecaption{Summary of the sample.
\label{tab:samples}}
\tablehead{
\colhead{} & \multicolumn{7}{c}{{\it all} sample} & & \multicolumn{7}{c}{{\it bright} subsample} \\ \cline{2-8} \cline{10-16}
\colhead{} & \multicolumn{3}{c}{UD-COSMOS} & & \multicolumn{3}{c}{UD-SXDS} & & \multicolumn{3}{c}{UD-COSMOS} & & \multicolumn{3}{c}{UD-SXDS} \\ \cline{2-4} \cline{6-8} \cline{10-12} \cline{14-16}
\colhead{$z_{\rm LAE}$} & \colhead{$L_{\rm Ly\alpha}^{\rm min}$} & \colhead{$L_{\rm Ly\alpha}^{\rm mean}$} & \colhead{$N_{\rm LAE}$} & & \colhead{$L_{\rm Ly\alpha}^{\rm min}$} & \colhead{$L_{\rm Ly\alpha}^{\rm mean}$} & \colhead{$N_{\rm LAE}$} & & \colhead{$L_{\rm Ly\alpha}^{\rm min}$} & \colhead{$L_{\rm Ly\alpha}^{\rm mean}$} & \colhead{$N_{\rm LAE}$} & & \colhead{$L_{\rm Ly\alpha}^{\rm min}$} & \colhead{$L_{\rm Ly\alpha}^{\rm mean}$} & \colhead{$N_{\rm LAE}$} \\
\colhead{(1)} & \colhead{(2)} & \colhead{(3)} & \colhead{(4)} & & \colhead{(5)} & \colhead{(6)} & \colhead{(7)} & & \colhead{(8)} & \colhead{(9)} & \colhead{(10)} & & \colhead{(11)} & \colhead{(12)} & \colhead{(13)}
}
\startdata
2.2 & 42.2 & 42.5 &  289 & & ---  & ---  & --- & & 42.6 & 42.9 &  37 & & --- & --- & --- \\
3.3 & 42.1 & 42.5 &  762 & & ---  & ---  & --- & & 42.6 & 42.9 & 123 & & --- & --- & --- \\
5.7 & 42.6 & 42.8 &  210 & & 42.5 & 42.8 & 393 & & 42.7 & 42.9 & 125 & & 42.6 & 42.9 & 313 \\
6.6 & 42.8 & 43.0 &   56 & & 42.9 & 43.0 &  24 & & 42.8$^*$ & 43.0$^*$ & 56$^*$ & & 42.9$^*$ & 43.0$^*$ &  24$^*$ \\
Total &    &      & 1317 & &      &      & 464 & &      &      & 341 & &     &     & 376
\enddata
\tablecomments{Columns: (1) Redshift. (2)-(3) Minimum and mean Ly$\alpha$ luminosity of the {\it all} sample LAEs in the UD-COSMOS field, measured with a $2\arcsec$-diameter aperture and shown in units of log erg s$^{-1}$. (4) Number of the LAEs. The total number of the LAEs over $z=2.2-6.6$ are shown in the bottom row. (5)-(7) Same as Columns (2)-(4), but for the {\it all} subsample in the UD-SXDS field. (8)-(13) Same as Columns (2)-(7), but for the {\it bright} subsample.\\
$*$ At $z=6.6$, we treat the {\it all} sample also as the {\it bright} subsample in each field (see Section \ref{subsec:lae}).}
\end{deluxetable*}

\subsection{NonLAE Samples} \label{subsec:nonlae}

Although the intensity mapping technique can remove spurious signals from low-redshift interlopers, other systematics, such as the sky background and PSF, may still contaminate LAE signals.
To estimate the contribution from these systematics, we use foreground sources, which we hereafter term ``NonLAEs.''
Because NonLAEs should correlate only with the systematics, but not with Ly$\alpha$ emission from LAEs, we estimate these systematics by applying the intensity mapping technique to NonLAEs.

\begin{figure}
\epsscale{1.1}
\plotone{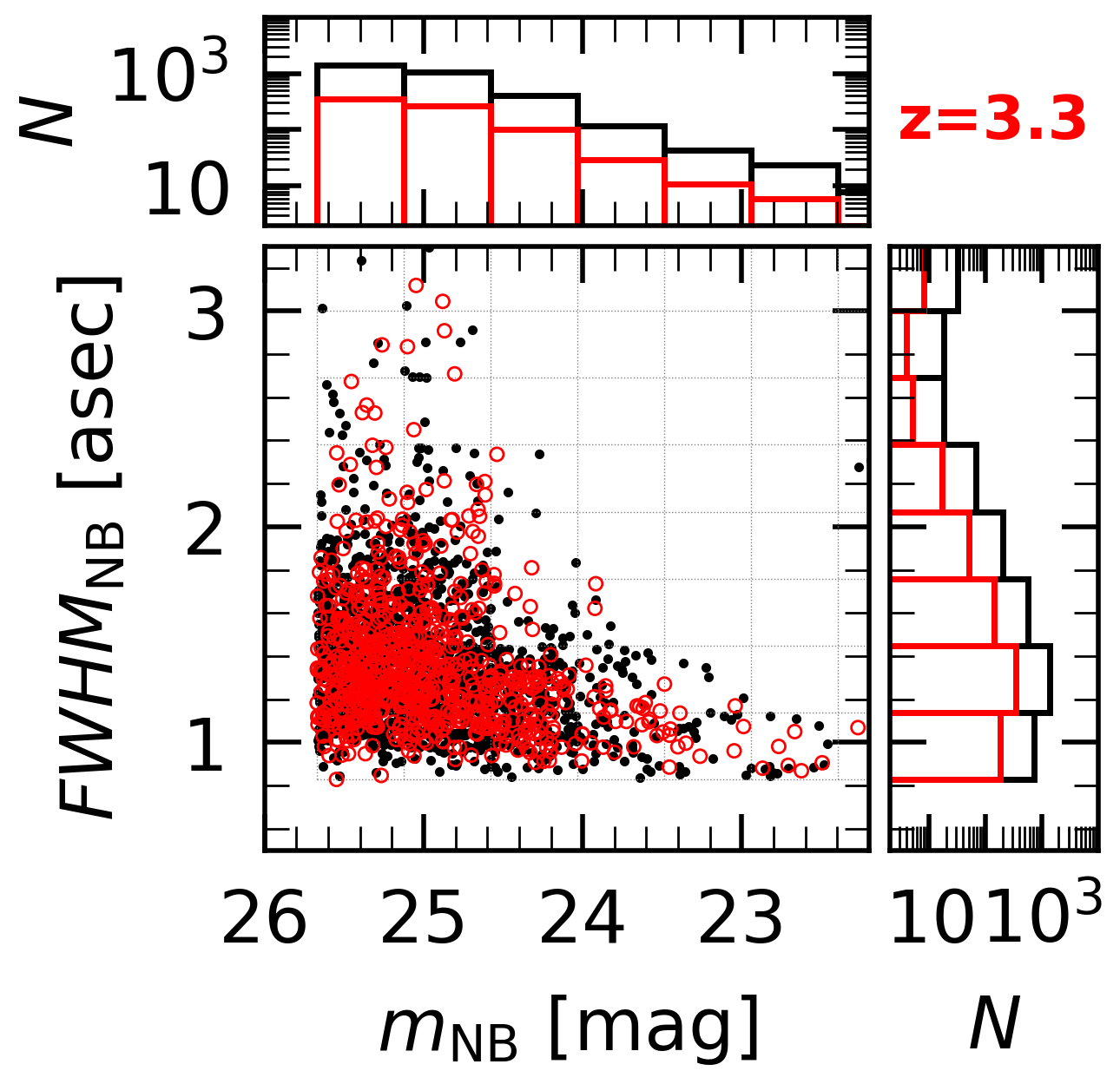}
\caption{{\bf Center:} NB FWHM (${\rm FWHM_{NB}}$) and magnitude ($m_{\rm NB}$) distributions of the {\it all} sample at $z=3.3$ (red circles), and the corresponding NonLAEs (black dots). {\bf Top:} $m_{\rm NB}$ histogram of the LAEs and NonLAEs (red and black bars, respectively). {\bf Right:} Same as the top panel, but for ${\rm FWHM_{NB}}$.}
\label{fig:2dhist}
\end{figure}

We construct NonLAE samples as follows.
First, we detect sources in the NB images using {\tt SExtractor} \citep{Bertin-Arnouts96}.
Second, we select only sources that are sufficiently bright ($\lesssim26$ mag) in the $g$-, $r2$-, and $i2$-bands, to remove spurious sources and artifacts.
Third, we randomly select the sources such that they have the same sky, FWHM, and $m_{\rm NB}$ distributions as those of the LAEs in each field at each redshift.
In this way, $\sim10^3$ sources are selected, which we define as NonLAEs (see Figure \ref{fig:2dhist} for the FWHM-$m_{\rm NB}$ distributions of the $z=3.3$ LAEs and corresponding NonLAEs).
We note that only $<1$ \% of the NonLAEs meet the color selection criteria of LAEs defined by \citet[see their Section 2]{Ono+21}.

\section{Intensity Mapping Analysis} \label{sec:im}

In this section, we derive the SB radial profiles of Ly$\alpha$ emission around the LAEs using the intensity mapping technique.

\subsection{Cross-correlation Functions} \label{subsec:sb}

We compute SB as a cross-correlation function between given band (XB) emission intensities and given objects (OBJs), ${\rm SB_{XB\times OBJ}}$\footnote{The right-hand side of Equation \ref{eqn:sb} represents a cross-correlation function, which has usually been referred to as $\xi$ in previous work \citep[e.g.][]{Croft+16,Croft+18,Bielby+17,Momose+21a,Momose+21b}. However, we refer to this function as `SB,' since the cross-correlation function is equivalent to surface brightness in our analyses.}, via
\begin{equation}
{\rm SB_{XB\times OBJ,\nu}}(r)=\frac{1}{N_{r,{\rm OBJ}}}\sum_{i=1}^{N_{r,{\rm OBJ}}}\mu^{(\rm XB)}_{\nu,i} \label{eqn:sb}
\end{equation}
and
\begin{equation}
{\rm SB_{XB\times OBJ}}(r)={\rm SB_{XB\times OBJ,\nu}}(r)\times{\rm FWHM_{XB}}.
\end{equation}
The pixel of the $i$-th pixel-OBJ pair has a pixel value of $\mu^{(\rm XB)}_{\nu,i}$ in the XB image in units of erg s$^{-1}$ cm$^{-2}$ Hz$^{-1}$ arcsec$^{-2}$.
In NB387, we multiply $\mu^{(\rm NB387)}_{\nu,i}$ by 1.5 to correct fot the zero-point offset of 0.45 mag (see Section \ref{subsec:lae}).
The summation runs over the $N_{r,{\rm OBJ}}$ pixel-OBJ pairs that are separated by a spatial distance $r$.
We use pixels at distances of between $1\farcs5$ and $40\arcsec$ from each OBJ (corresponding to the outer part of the CGM and outside), which were then divided into six radial bins.
A total of $\sim(1-2)\times10^9$ pixels were used for the calculation at each redshift.
${\rm FWHM_{XB}}$ represents the XB filter width (in units of Hz) corrected for IGM attenuation, derived via
\begin{equation}
{\rm FWHM_{XB}}=\frac{\int_0^\infty e^{-\tau_{\rm eff}(\nu)}T_{\rm XB}(\nu){\rm d}\nu/\nu}{T_{\rm XB}(\nu_\alpha)/\nu_\alpha},
\end{equation}
where $T_{\rm XB}(\nu)$ denotes the transmittance of the XB filter, and $\nu_\alpha$ is the the observed Ly$\alpha$ line frequency.
We adopt the IGM optical depth $\tau_{\rm eff}(\nu)$ from \citet{Inoue+14}.

The statistical uncertainty of ${\rm SB_{XB\times OBJ}}$ is estimated by the bootstrap method.
We randomly resample objects while keeping the sample size and calculated ${\rm SB_{XB\times OBJ}}$.
We then repeat the resampling $10^4$ times, adopting the $1\sigma$ standard deviation of the ${\rm SB_{XB\times OBJ}}$ values as the $1\sigma$ statistical uncertainty of the original ${\rm SB_{XB\times OBJ}}$.

\subsection{Ly$\alpha$ Surface Brightness} \label{subsec:sb_lya}

\begin{figure}
\epsscale{1.1}
\plotone{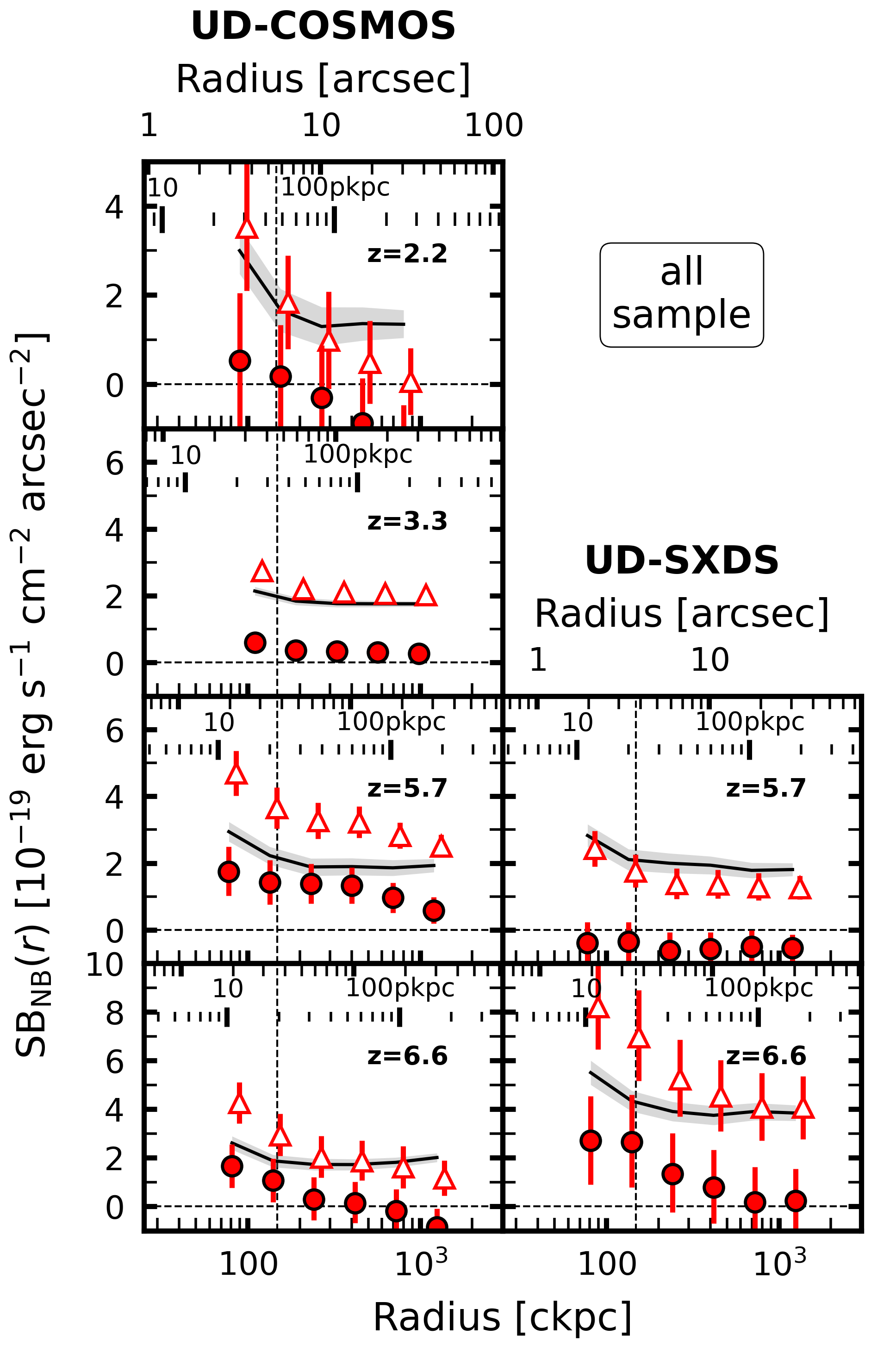}
\caption{SB radial profiles of the {\it all} sample. {\bf Left:} ${\rm SB_{NB}}$ in the UD-COSMOS field at $z=2.2$, 3.3, 5.7, and 6.6, from top to bottom. The red triangles and circles show SB before and after systematics subtraction (${\rm SB_{NB\times LAE}}$ and ${\rm SB_{NB}}$), respectively. The black solid lines represent the systematics that are estimated with the NonLAEs (${\rm SB_{NB\times NonLAE}}$). The red error bars and gray shades are the $1\sigma$ uncertainties estimated by the bootstrap method. The vertical black dashed line represents $R_{\rm vir}$ of a DMH with $M_{\rm halo}=10^{11}$ $M_\odot$, while the horizontal black line represents ${\rm SB}=0$. The upper, middle, and lower horizontal axes in each panel show radii in units of arcsec, pkpc, and ckpc, respectively. The data points of ${\rm SB_{NB\times LAE}}$ are slightly shifted along the horizontal axis for clarity. {\bf Right:} Same as the left column, but in the UD-SXDS field at $z=5.7$ and 6.6 in the top and bottom panels, respectively. The S/N ratios of ${\rm SB_{NB}}$ are 4.1, 1.6, and 2.2 at $z=3.3$, 5.7, and 6.6, respectively (they are based on ${\rm SB_{NB}}$ averaged over the UD-COSMOS and UD-SXDS fields, at $z=5.7$ and 6.6). There is no clear detection at $z=2.2$.}
\label{fig:sb_nb_all}
\end{figure}

\begin{figure}
\epsscale{1.1}
\plotone{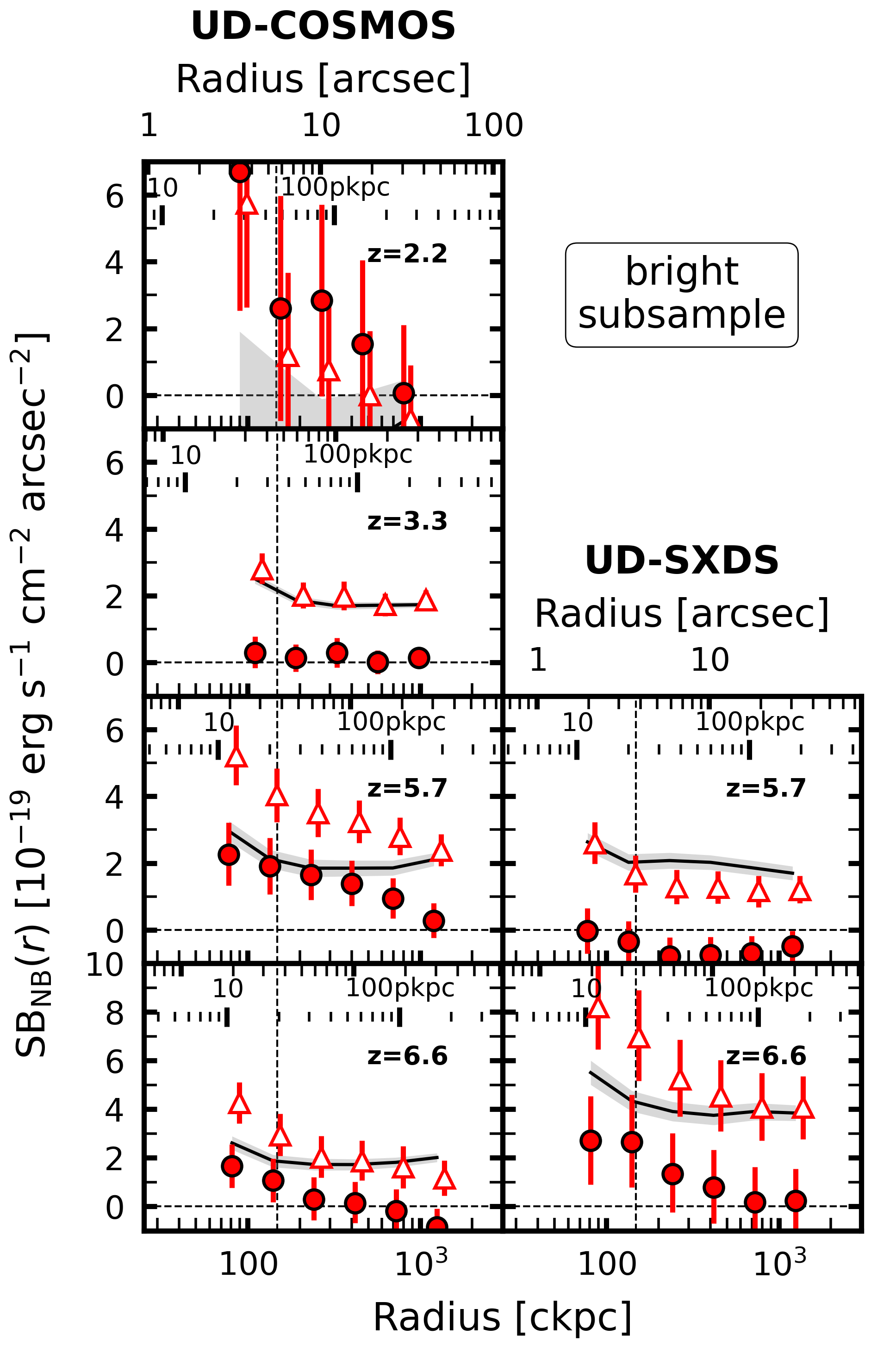}
\caption{Same as Figure \ref{fig:sb_nb_all}, but for the {\it bright} subsample. The S/N ratios of ${\rm SB_{NB}}$ are 1.5 and 1.4 at $z=2.2$ and 5.7, respectively, while there is no clear detection at $z=3.3$. The results at $z=6.6$ (bottom panels) are the same as those in Figure \ref{fig:sb_nb_all} since we treat the {\it all} sample also as the {\it bright} subsample at $z=6.6$.}
\label{fig:sb_nb_bright}
\end{figure}

We estimate the SB of Ly$\alpha$ emission (${\rm SB_{Ly\alpha}}$) around the LAEs as follows.
First, we subtract the systematics (${\rm SB_{NB\times NonLAE}}$) from the emission from the LAEs (${\rm SB_{NB\times LAE}}$) via
\begin{equation}
{\rm SB_{NB}=SB_{NB\times LAE}-SB_{NB\times NonLAE}}. \label{eqn:sb_nb_halo}
\end{equation}
Uncertainties in ${\rm SB_{NB}}$ propagate from those of ${\rm SB_{NB\times LAE}}$ and ${\rm SB_{NB\times NonLAE}}$.
We present the radial profiles of the ${\rm SB_{NB\times LAE}}$, ${\rm SB_{NB\times NonLAE}}$, and ${\rm SB_{NB}}$ of the {\it all} sample and {\it bright} subsample in Figures \ref{fig:sb_nb_all} and \ref{fig:sb_nb_bright}, respectively.

We use Fisher's method \citep{Fisher70} to estimate the S/N ratios of ${\rm SB_{NB}}$ over all the radial bins, following \citet{Kakuma+21}. Generally, a $p$-value is expressed as
\begin{equation}
p=\int_{{\rm S/N}}^\infty\mathcal{N}(x;\mu=0,\sigma=1){\rm d}x, \label{eqn:sn_to_p}
\end{equation}
where $\mathcal{N}(x;\mu=0,\sigma=1)$ is a Gaussian distribution with an expected value $\mu=0$ and a variance $\sigma^2=1$.
We thus use this equation to convert the S/N ratio in the $i$-th radial bin (S/N$_i$) into the $p$-value in that bin ($p_i$).
The $\chi^2$ value over all the radial bins ($1\leq i\leq N$), $\hat{\chi}^2$, is then calculated as $\hat{\chi}^2=-2\sum_{i=1}^N\ln(p_i)$.
Since $\hat{\chi}^2$ follows a $\chi^2_{2N}$ distribution with $2N$ degrees of freedom, $\chi^2(x;{\rm dof}=2N)$, the $p$-value over all the radial bins, $\hat{p}$, is derived as
\begin{equation}
\hat{p}=\int_{\hat{\chi}^2}^\infty\chi^2(x;{\rm dof}=2N){\rm d}x. \label{eqn:chi2_to_p}
\end{equation}
We convert this to the S/N ratio over all the radial bins by solving Equation (\ref{eqn:sn_to_p}) for S/N.
We use the radial bins at $<1$ cMpc.

The black vertical dashed lines in Figures \ref{fig:sb_nb_all} and \ref{fig:sb_nb_bright} indicate $R_{\rm vir}$.
Following the observational results by \citet{Ouchi+10} and \citet{Kusakabe+18}, we assume that the DMHs hosting LAEs have halo masses ($M_{\rm halo}$) of $10^{11}\ M_\odot$ at all the redshifts, which corresponds to a $R_{\rm vir}$ value of $\sim150$ ckpc.

As presented in Figures \ref{fig:sb_nb_all} and \ref{fig:sb_nb_bright}, at $z=2.2$, we see no clear detection for the {\it all} sample, although ${\rm SB_{NB\times LAE}}$ slightly exceeds ${\rm SB_{NB\times NonLAE}}$ for the {\it bright} subsample with ${\rm S/N}=1.5$.
At $z=3.3$, ${\rm SB_{NB}}$ of the {\it all} sample is significantly positive over wide scales from $\sim100$ ckpc to 1 cMpc with ${\rm S/N}=4.1$.
At $z=5.7$, ${\rm SB_{NB}}$ is significantly positive at $\sim80-10^3$ ckpc for the {\it all} sample and {\it bright} subsample in the UD-COSMOS field with ${\rm S/N}=4.6$ and 4.1, respectively.
Averaging ${\rm SB_{NB}}$ of the {\it all} sample in both fields with weights of the number of the LAEs in each field, we identify a tentatively positive signal with ${\rm S/N}=1.6$.
At $z=6.6$, ${\rm SB_{NB}}$ are positive at $\sim80-200$ ckpc in both fields.
Averaging ${\rm SB_{NB}}$ over the two fields, we tentatively identify a positive signal with ${\rm S/N}=2.3$.
We find that the emission SB at $z=3.3-6.6$ is as diffuse as $\sim10^{-20}-10^{-19}$ erg s$^{-1}$ cm$^{-2}$ arcsec$^{-2}$.
We note that ${\rm SB_{NB\times LAE}}$ in the UD-SXDS field are systematically lower than ${\rm SB_{NB\times NonLAE}}$, as reported in \citet{Kakuma+21}.

\begin{figure}
\epsscale{1.1}
\plotone{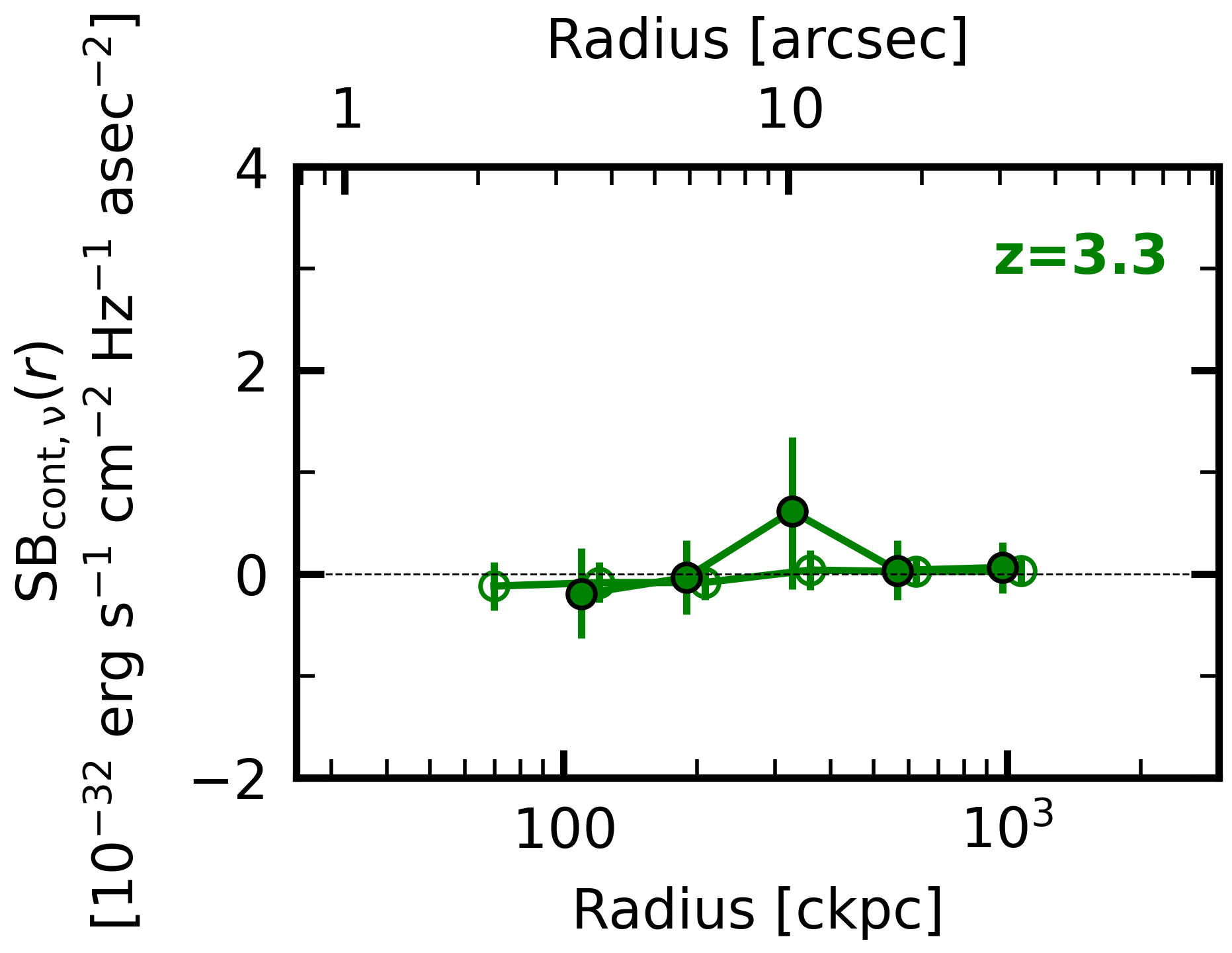}
\caption{Radial profiles of the UV continuum SB of the {\it all} sample (open) and {\it bright} subsample (filled) at $z=3.3$. The profiles are averages weighted with the number of the LAEs in the UD-COSMOS and UD-SXDS fields. The data points for {\it bright} subsample are slightly shifted along the horizontal axis for clarity. The data points should be treated as upper limits, since ${\rm SB_{cont,\nu}}$ neglects signals from the PSF. ${\rm SB_{cont,\nu}}$ agree with null detection within $1\sigma$ uncertainties for both sample, which disfavors the UV continuum contribution to ${\rm SB_{NB}}$.}
\label{fig:sb_cont_all}
\end{figure}

UV continuum emission might contribute to the ${\rm SB_{NB}}$ in addition to the Ly$\alpha$ line emission.
We thus estimate SB of the UV continuum emission, ${\rm SB_{cont,\nu}}$, using
\begin{equation}
{\rm SB_{cont,\nu}<SB_{BB,\nu}\equiv SB_{BB\times LAE,\nu}-SB_{BB\times random,\nu}}, \label{eqn:sb_bb_halo}
\end{equation}
where ${\rm SB_{BB\times LAE,\nu}}$ (${\rm SB_{BB\times random,\nu}}$) represents the SB value that is derived from the cross-correlation function between the BB images and LAEs (random sources) in units of erg s$^{-1}$ cm$^{-2}$ Hz$^{-1}$ arcsec$^{-2}$.
To estimate the sky background, we use random sources, not NonLAEs, since it is difficult to match the ${\rm FWHM_{NB}}$-$m_{\rm NB}$ distributions of NonLAEs with those of the LAEs due to faintness of the LAEs in the BB.
Since ${\rm SB_{BB\times random,\nu}}$ neglects neglects signals from the PSF, ${\rm SB_{BB,\nu}}$ should be treated as the upper limit of ${\rm SB_{cont,\nu}}$.
As shown in Figure \ref{fig:sb_cont_all} as an example, the values of ${\rm SB_{cont,\nu}}$ at $z=2.2-6.6$ are consistent with null detection within $\sim(1-2)\sigma$ uncertainties.
Additionally, we confirm that the UV continuum emission contributing to ${\rm SB_{NB}}$, i.e., ${\rm SB_{cont,\nu}}\times{\rm FWHM_{NB}}$, is negligible compared to ${\rm SB_{NB}}$.
Therefore, we hereafter assume that ${\rm SB_{NB}}$ is equivalent to the Ly$\alpha$ SB (${\rm SB_{Ly\alpha}}$).

In summary, we identify very diffuse ($\sim10^{-20}-10^{-19}$ erg s$^{-1}$ cm$^{-2}$ arcsec$^{-2}$) Ly$\alpha$ signals beyond $R_{\rm vir}$ around the {\it all} sample LAEs at $z=3.3$ at the $4.1\sigma$ level.
We also potentially detect positive signals around the {\it all} sample LAEs at $z=5.7$ and 6.6 and {\it bright} subsample LAEs with ${\rm S/N}\sim2$.
These results imply the potential existence of very diffuse and extended Ly$\alpha$ emission around $z=2.2-6.6$ LAEs.

Again, a major update compared to \citet{Kakuma+21} is that we apply the intensity mapping technique newly to the CHORUS data to investigate extended Ly$\alpha$ emission at $z=2.2-3.3$.
In particular, we identify extended Ly$\alpha$ emission around the $z=3.3$ LAEs with the S/N levels comparable or higher than those at $z=5.7$ and 6.6, probably thanks to the large number of LAEs at $z=3.3$ ($N_{\rm LAE}=762$).
Our results at $z=5.7$ and 6.6 are consistent with those obtained in \citet{Kakuma+21}, which is as expected from similar properties of the LAEs of \citet{Kakuma+21} and our {\it all} sample, such as $N_{\rm LAE}$, the $L_{\rm Lya}$ ranges, and the sky distributions (Section \ref{subsec:lae}).
In the next section, we compare our results with previous work including \citet{Kakuma+21} in detail.

\subsection{Comparison with Previous Work} \label{subsec:compare_obs}

\begin{figure*}
\epsscale{1.1}
\plotone{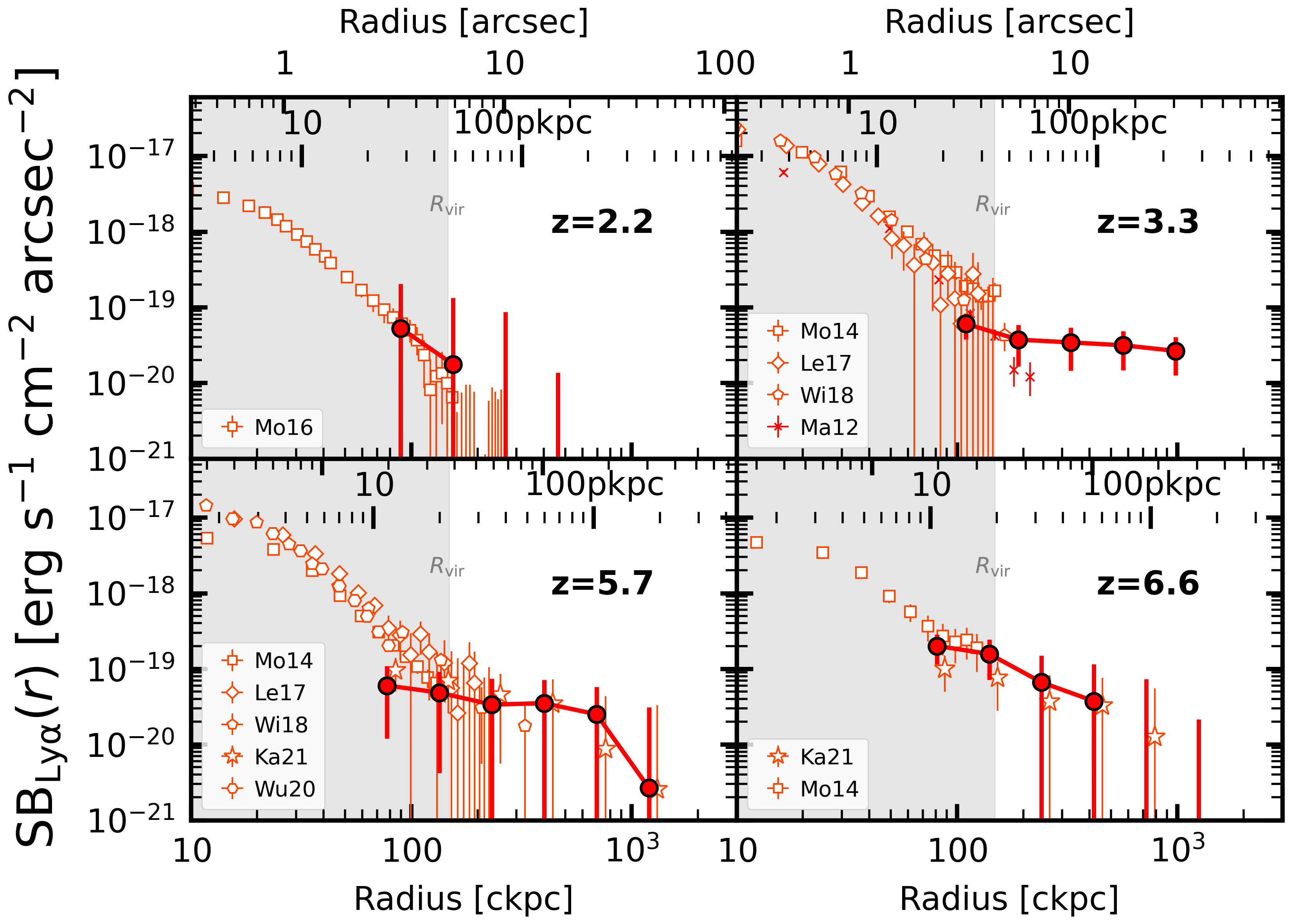}
\caption{Comparison of the ${\rm SB_{Ly\alpha}}$ radial profiles at $z=2.2$ (top left), 3.3 (top right), 5.7 (bottom left), and 6.6 (bottom right). All the ${\rm SB_{Ly\alpha}}$ profiles are corrected by $(1+z)^{-4}$ to match to each redshift. The gray shades illustrate the regions whose radius is smaller than $R_{\rm vir}$ of a DMH with $M_{\rm halo}=10^{11}\ M_\odot$. The filled red circles show the ${\rm SB_{Ly\alpha}}$ profiles of our {\it all} sample. The profiles at $z=5.7$ and 6.6 are averaged over the UD-COSMOS and UD-SXDS fields. The other red symbols represent the profiles taken from previous studies: \citet[Ma12: crosses]{Matsuda+12}, \citet[Mo14: squares]{Momose+14}, \citet[Mo16: squares]{Momose+16}, \citet[Le17: diamonds]{Leclercq+17}, \citet[Wi18: pentagons]{Wisotzki+18}, \citet[Wu20: hexagons]{Wu+20}, and \citet[Ka21: stars]{Kakuma+21}. See Table \ref{tab:compare_obs} for details of these samples. We omit the data points below the detection limits defined in the literature. The ${\rm SB_{Ly\alpha}}$ profiles taken from \citet{Leclercq+17}, \citet{Wisotzki+18}, and \citet{Wu+20} are normalized such that $L_{\rm Ly\alpha}$ values measured in a central $2\arcsec$ diameter equal to those of our {\it all} sample at each redshift. Some data points are slightly shifted along the horizontal axes for clarity. The ${\rm SB_{Ly\alpha}}$ profiles of our {\it all} sample are in good agreements with those taken from the literature.}
\label{fig:sb_lya_vsobs_all}
\end{figure*}

We compare our ${\rm SB_{Ly\alpha}}$ radial profiles with those of previous studies at $z\sim2.2$, 3.3, 5.7, and 6.6 (Figure \ref{fig:sb_lya_vsobs_all}).
We compile the data taken from \citet{Momose+14}, \citet{Momose+16}, \citet{Leclercq+17}, \citet{Wisotzki+18}, \citet{Wu+20}, and \citet{Kakuma+21}, which are summarized in Table \ref{tab:compare_obs}.
Because SB is affected by the cosmological dimming effect, all the ${\rm SB_{Ly\alpha}}$ profiles, including ours, are scaled by $(1+z)^{-4}$ to $z=2.2$, 3.3, 5.7, and 6.6 in each panel.
We also shift the radii of the ${\rm SB_{Ly\alpha}}$ profiles in units of cpkc by $(1+z)$, while fixing the radii in pkpc.
For our samples at $z=5.7$ and 6.6, we hereafter present ${\rm SB_{Ly\alpha}}$ averaged at each redshift over the UD-COSMOS and UD-SXDS fields weighting by the number of LAEs in each field, unless otherwise stated.

Although the ${\rm SB_{Ly\alpha}}$ profiles are measured under different seeing sizes, the typical image PSF FWHMs are as small as $\lesssim1\farcs5$, corresponding to $\lesssim40-60$ ckpc at $z=2-7$ \citep[e.g.][]{Momose+14,Ono+21}.
Since we focus on ${\rm SB_{Ly\alpha}}$ profiles at larger scales of $\gtrsim100$ ckpc, PSF differences are unlikely to affect the following discussion.

\citet{Momose+16} found that ${\rm SB_{Ly\alpha}}$ profiles depend on $L_{\rm Ly\alpha}$ of the galaxy.
To avoid this dependency, we take the data from \citet{Momose+14}, \citet{Momose+16}, and \citet{Kakuma+21}, because their LAEs have $L_{\rm Ly\alpha}$ values similar to those of our {\it all} samples in the same $2\arcsec$-diameter aperture size.
We additionally take the data from \citet{Leclercq+17}, \citet{Wisotzki+18}, and \citet{Wu+20}, but these samples have different $L_{\rm Ly\alpha}$ values measured in different aperture sizes.
Therefore, for precise comparisons, we normalize the ${\rm SB_{Ly\alpha}}$ profiles of these samples such that ${\rm SB_{Ly\alpha}}$ integrated over a central $2\arcsec$-diameter aperture ($=4\pi d_{\rm L}^2\int_{0''}^{1''}{\rm SB_{Ly\alpha}}(r)\cdot2\pi r{\rm d}r$) becomes equal to $L_{\rm Ly\alpha}$ of our {\it all} sample at each redshift.

\begin{deluxetable*}{lcccc}
\tablecaption{Summary of the observational studies used for comparison.
\label{tab:compare_obs}}
\tablehead{\colhead{Reference} & \colhead{$z$} & \colhead{$L_{\rm Ly\alpha}$} & \colhead{Sample} & \colhead{Method}}
\colnumbers
\startdata
$z\sim2.2$ & & & & \\
This Work ({\it all} sample) & 2.2 & 42.5 & 289 LAEs & intensity mapping (mean) \\
\citet{Momose+16} & 2.2 & 42.6 & 710 LAEs ($L_{\rm Ly\alpha}\geq10^{42.4}$ erg s$^{-1}$) & stacking (mean) \\ \tableline
$z\sim3.3$ & & & & \\
This Work ({\it all} sample) & 3.3 & 42.5 & 762 LAEs & intensity mapping (mean) \\
\citet{Momose+14} & 3.1 & 42.7 & 316 LAEs & stacking (mean) \\
\citet{Leclercq+17} & 3.28 & [42.5] & LAE MUSE\#106$^*$ & individual detection \\
\citet{Wisotzki+18} & $3-4$ & [42.5] & 18 LAEs ($L_{\rm Ly\alpha}>10^{42}$ erg s$^{-1}$) & stacking (median) \\
\citet{Matsuda+12} & 3.1 & --- & 894 LAEs ($26<BV<27$)$^\dagger$ & stacking (median) \\ \tableline
$z\sim5.7$ & & & & \\
This Work ({\it all} sample) & 5.7 & 42.8 & 650 LAEs & intensity mapping (mean) \\
\citet{Momose+14} & 5.7 & 42.7 & 397 LAEs & stacking (mean) \\
\citet{Leclercq+17} & 5.98 & [42.8] & LAE MUSE\#547$^*$ & individual detection \\
\citet{Wisotzki+18} & $5-6$ & [42.8] & 6 LAEs ($L_{\rm Ly\alpha}>10^{42}$ erg s$^{-1}$) & stacking (median) \\
\citet{Wu+20} & 5.7 & [42.8] & 310 LAEs & stacking (median) \\
\citet{Kakuma+21} & 5.7 & 42.9 & 425 LAEs & intensity mapping (mean) \\ \tableline
$z\sim6.6$ & & & & \\
This Work ({\it all} sample) & 6.6 & 43.0 & 80 LAEs & intensity mapping (mean) \\
\citet{Momose+14} & 6.6 & 42.7 & 119 LAEs & stacking (mean) \\
\citet{Kakuma+21} & 6.6 & 42.8 & 396 LAEs & intensity mapping (mean)
\enddata
\tablecomments{Columns: (1) Reference. (2) Redshift. (3) Mean or median Ly$\alpha$ luminosity of the sample within a $2\arcsec$-diameter aperture in units of log erg s$^{-1}$. We normalize the ${\rm SB_{Ly\alpha}}$ profiles of \citet{Leclercq+17}, \citet{Wisotzki+18}, and \citet{Wu+20} such that ${\rm SB_{Ly\alpha}}$ integrated over a central $2\arcsec$-diameter aperture becomes equal to $L_{\rm Ly\alpha}$ of our {\it all} sample at each redshift, which are indicated by the brackets. (4) Sample used for comparison. The parentheses indicate specific subsamples. (5) Method for deriving the ${\rm SB_{Ly\alpha}}$ profiles.\\
$^*$ ID of the individual LAE. See Section \ref{subsec:compare_obs} for the variance among the individual LAEs.\\
$^\dagger$ $BV\equiv(2B+V)/3$, where $B$ and $V$ are $B$- and $V$-band magnitudes.
}
\end{deluxetable*}

The top left panel in Figure \ref{fig:sb_lya_vsobs_all} presents ${\rm SB_{Ly\alpha}}$ radial profiles at $z=2.2$.
We compare the results of \citet[$L_{\rm Ly\alpha}=10^{42.6}$ erg s$^{-1}$ subsample]{Momose+16} against our {\it all} sample.
We found that the ${\rm SB_{Ly\alpha}}$ profile of \citet{Momose+16} is in good agreement with that of our {\it all} sample at $r\sim100$ ckpc.

In the top right panel of Figure \ref{fig:sb_lya_vsobs_all}, we show the ${\rm SB_{Ly\alpha}}$ profiles at $z=3.3$.
We compare the results of \citet[$z=3.1$ LAEs]{Momose+14}, \citet[an individual LAE MUSE\#106]{Leclercq+17}, and \citet[$L_{\rm Ly\alpha}>10^{42}$ erg s$^{-1}$ subsample at $z=3-4$]{Wisotzki+18}, and our {\it all} sample.
The ${\rm SB_{Ly\alpha}}$ profiles from the literature approximately agree with that of our {\it all} sample even at $r\sim R_{\rm vir}$.
We additionally show the results from a subsample of \citet{Matsuda+12} with continuum magnitudes $BV\equiv(2B+V)/3$ of $26<BV<27$ (typical value range for LAEs), where $B$ and $V$ are $B$- and $V$-band magnitudes measured with the Subaru/SC.
The ${\rm SB_{Ly\alpha}}$ profile of \citet{Matsuda+12} also agrees with that of our {\it all} sample around $r\sim R_{\rm vir}$.
We note that, although the results of \citet{Leclercq+17} are represented by their individual LAE MUSE\#6905, their LAEs have similar ${\rm SB_{Ly\alpha}}$ profiles when the amplitudes are normalized to match the $L_{\rm Ly\alpha}$ values at $r\leq1\arcsec$.

The ${\rm SB_{Ly\alpha}}$ profiles at $z=5.7$ are displayed in the bottom left panel of Figure \ref{fig:sb_lya_vsobs_all}.
We compare the results of \citet[$z=5.7$ LAEs]{Momose+14}, \citet[an individual LAE MUSE\#547]{Leclercq+17}, \citet[$L_{\rm Ly\alpha}>10^{42}$ erg s$^{-1}$ subsample at $z=5-6$]{Wisotzki+18}, \citet[$z=5.7$ LAEs]{Wu+20}, \citet[$z=5.7$ LAEs]{Kakuma+21}, and our {\it all} sample.
The ${\rm SB_{Ly\alpha}}$ profile of our {\it all} sample agrees well with those of \citet{Momose+14}, \citet{Leclercq+17}, \citet{Wisotzki+18} and \citet{Wu+20} at $r\sim80-200$ ckpc, and with that of \citet{Kakuma+21} up to $r\sim1$ cMpc.

The bottom right panel of Figure \ref{fig:sb_lya_vsobs_all} shows ${\rm SB_{Ly\alpha}}$ profiles at $z=6.6$ taken from \citet[$z=6.6$ LAEs]{Momose+14}, \citet[$z=6.6$ LAEs]{Kakuma+21}, and our {\it all} sample.
The ${\rm SB_{Ly\alpha}}$ profile of our {\it all} sample is consistent with those of \citet{Momose+14} and \citet{Kakuma+21} up to the scales of $r\sim100$ ckpc and 1 cMpc, respectively.

In summary, our ${\rm SB_{Ly\alpha}}$ profiles are in good agreement with those of the previous studies at each redshift, provided that the LAEs have similar $L_{\rm Ly\alpha}$ values at $r\leq1\arcsec$.
Our ${\rm SB_{Ly\alpha}}$ profiles ($r\gtrsim80$ ckpc) are smoothly connected with the inner ($r\lesssim100$ ckpc) profiles taken from the literature at $\sim100$ ckpc, and extend to larger scales.

\begin{figure}
\epsscale{1.1}
\plotone{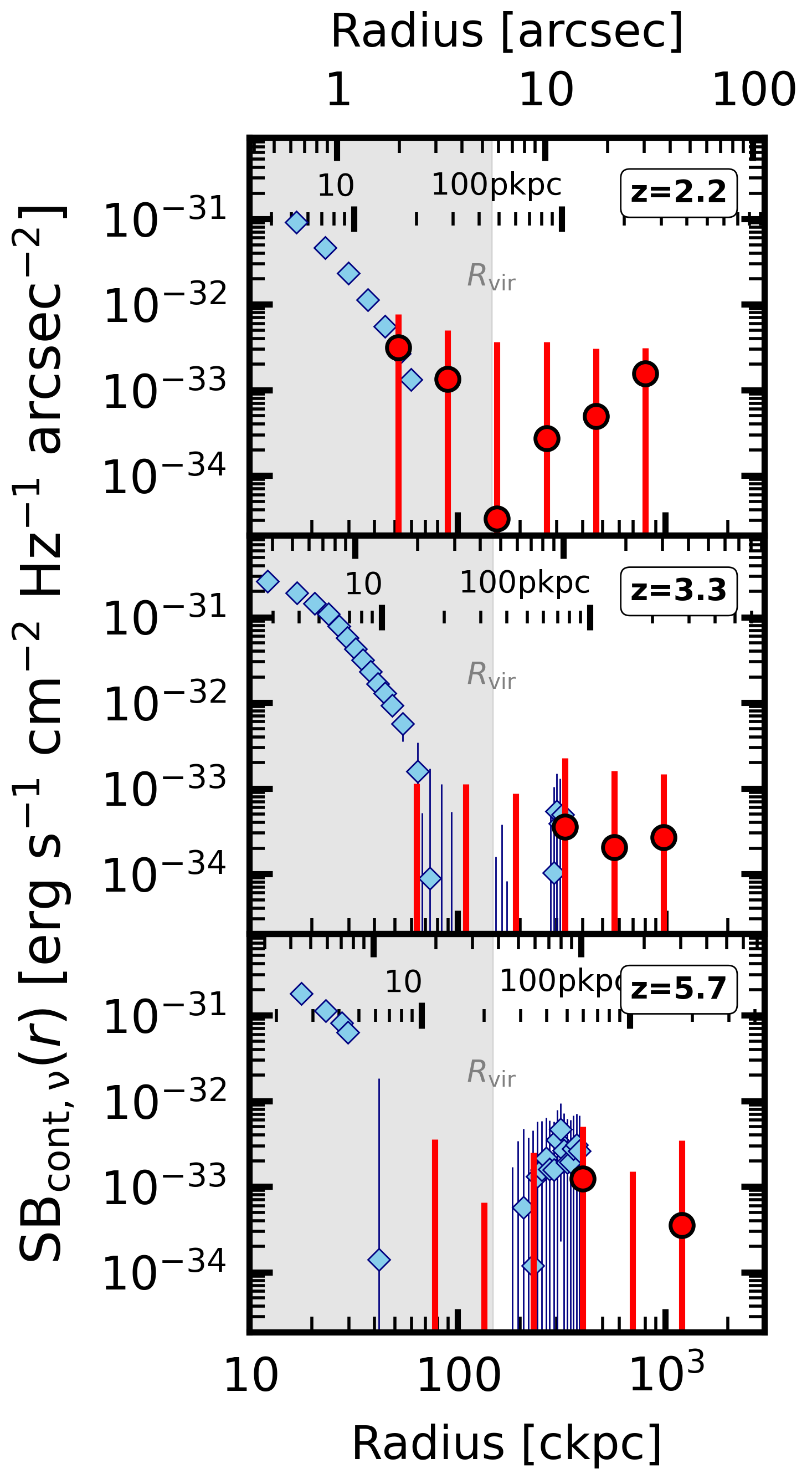}
\caption{Comparison of the ${\rm SB_{cont,\nu}}$ radial profiles at $z=2.2$ (top), 3.3 (middle), and 5.7 (bottom). The red circles and cyan diamonds show the ${\rm SB_{cont,\nu}}$ profiles of our {\it all} sample and \citet{Momose+14,Momose+16}, respectively (we use the data of \citealt{Momose+16} at $z=2.2$, and of $z=3.1$ and 5.7 LAEs of \citealt{Momose+14} at $z=3.3$ and 5.7, respectively). The gray shades illustrate the regions inside $R_{\rm vir}$ of a DMH with $M_{\rm halo}=10^{11}\ M_\odot$. The ${\rm SB_{cont,\nu}}$ profiles of our {\it all} sample are in good agreements with those of \citet{Momose+14,Momose+16}.}
\label{fig:sb_cont_vsobs_all}
\end{figure}

In Figure \ref{fig:sb_cont_vsobs_all}, we compare the ${\rm SB_{cont,\nu}}$ radial profiles between \citet{Momose+14,Momose+16} and our {\it all} sample at $z=2.2$, 3.3, and 5.7 (the data of \citealt{Momose+14,Momose+16} are of the same LAEs as used in Figure \ref{fig:sb_lya_vsobs_all}).
We find that the ${\rm SB_{cont,\nu}}$ profiles roughly agree at $r\lesssim400$ ckpc, although the uncertainties are large.
The ${\rm SB_{cont,\nu}}$ profiles are much less extended than ${\rm SB_{Ly\alpha}}$ profiles, which was also suggested by \citet{Momose+14}, \citet{Momose+16}, and \citet{Wu+20}.
We note the profiles at $z=6.6$ are not compared here because our sample is $\sim0.3$ dex brighter than that of \citet{Momose+14}.

\section{Discussion} \label{sec:discussion}

\subsection{Redshift Evolution of Extended Ly$\alpha$ Emission Profiles} \label{subsec:zevolution}

\begin{figure}
\epsscale{1.1}
\plotone{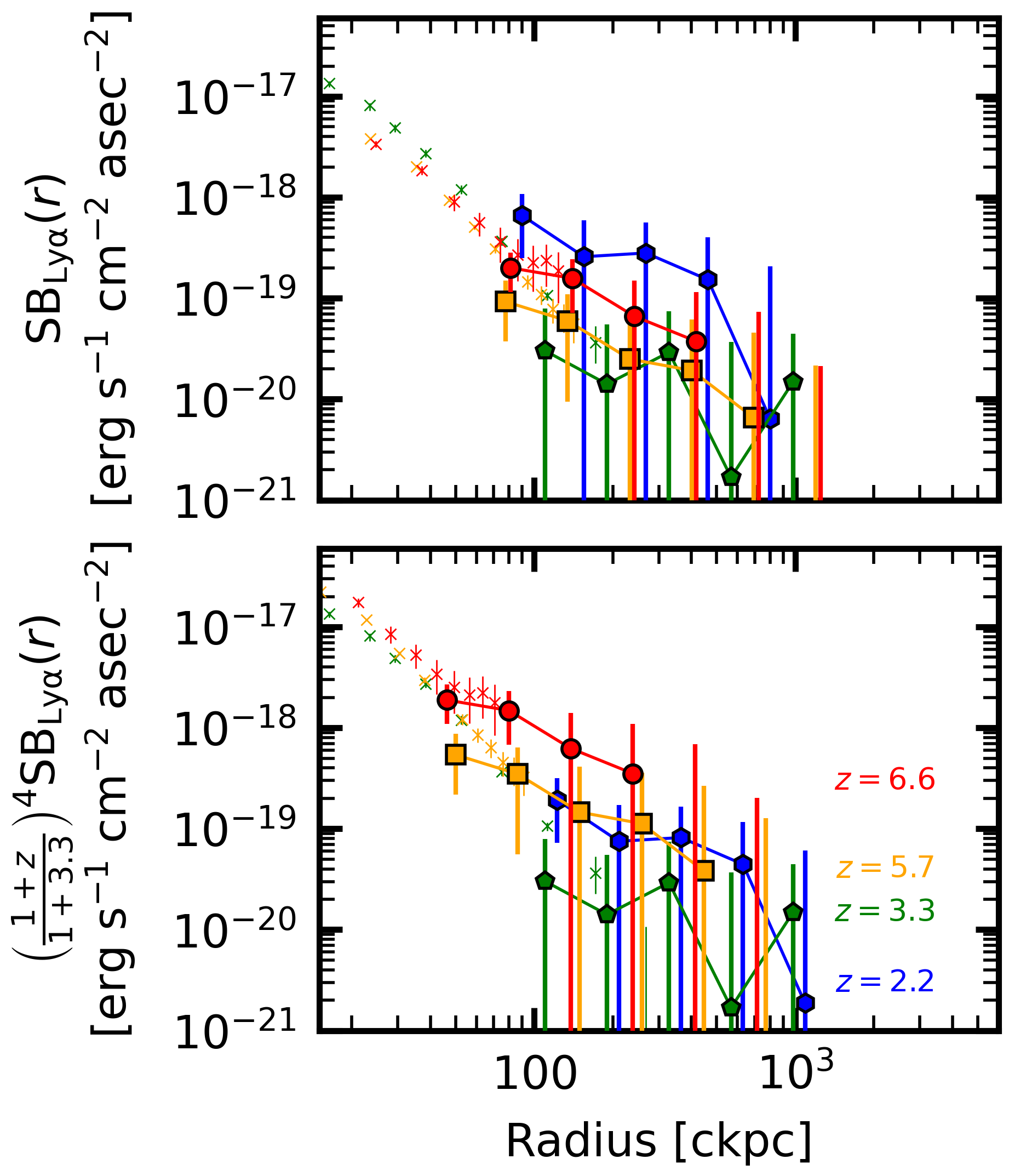}
\caption{Comparison between ${\rm SB_{Ly\alpha}}$ radial profiles at $z=2.2-6.6$. {\bf Top:} Observed ${\rm SB_{Ly\alpha}}$ profiles. The blue hexagons, green pentagons, orange squares, and red circles represent the profiles of our {\it bright} subsamples at $z=2.2$, 3.3, 5.7, and 6.6, respectively. The green, orange, and red crosses denote profiles taken from \citet{Wisotzki+18}, \citet{Momose+14}, and \citet{Momose+14}, which are normalized to $z=3.3$, 5.7, and 6.6, respectively. The ${\rm SB_{Ly\alpha}}$ profile of \citet{Wisotzki+18} is normalized such that $L_{\rm Lya}$ in $r\leq1\arcsec$ matches to that of our {\it all} sample (see Section \ref{subsec:compare_obs}). We again note that our profiles here are based on {\it bright} subsamples and thus differ from those shown in Figure \ref{fig:sb_lya_vsobs_all}. {\bf Bottom:} Same as the top panel, but showing the profiles corrected for the cosmological dimming effect, i.e., the intrinsic ${\rm SB_{Ly\alpha}}$ profiles. We tentatively identify an increasing trend roughly by $(1+z)^3$ toward high redshifts, albeit with the large uncertainties.}
\label{fig:sb_lya_zevolve_bright}
\end{figure}

In this section, we investigate the redshift evolution of ${\rm SB_{Ly\alpha}}$ profiles of extended Ly$\alpha$ emission.
In Figure \ref{fig:sb_lya_zevolve_bright}, we compare the ${\rm SB_{Ly\alpha}}$ profiles of our {\it bright} subsamples at $z=2.2-6.6$ as a function of radius in units of ckpc.
We also present the results taken from \citet[$z=5.7$ and 6.6 LAEs]{Momose+14} and \citet[$L_{\rm Ly\alpha}>10^{42}$ erg s$^{-1}$ subsample at $z=3-4$]{Wisotzki+18}.
Because our {\it bright} subsamples have uniform $L_{\rm Ly\alpha}$ values ($\sim10^{42.9}-10^{43.0}$ erg s$^{-1}$) over $z=2.2-6.6$, the $L_{\rm Ly\alpha}$ differences between the redshifts are unlikely to influence the following discussion.

The top panel of Figure \ref{fig:sb_lya_zevolve_bright} shows the observed ${\rm SB_{Ly\alpha}}$ profiles.
We identify no significant difference among the ${\rm SB_{Ly\alpha}}$ profiles beyond the $1\sigma$ uncertainties at $r\sim100-1000$ ckpc over $z=2.2-6.6$, while the uncertainties are large.
This finding is consistent with that of \citet{Kakuma+21} at $z=5.5-6.6$.
There is no significant difference also in the profiles at $r<100$ ckpc, which was also suggested by MUSE observations \citep[see also Figure 11 of \citealt{Byrohl+20}]{Leclercq+17} at $3<z<6$.

Observed ${\rm SB_{Ly\alpha}}$ profiles are affected by the cosmological dimming effect.
To correct this effect, we shift the observed ${\rm SB_{Ly\alpha}}$ profiles vertically by $(1+z)^4/(1+3.3)^4$ and horizontally by $(1+3.3)/(1+z)$, which are hereafter termed as the {\it intrinsic} profiles (we match the profiles to $z=3.3$ just for visibility).
The intrinsic profiles are presented in the bottom panel of Figure \ref{fig:sb_lya_zevolve_bright}.
There is a tentative increasing trend in the intrinsic profiles toward $z=6.6$, although those at $z=2.2$ and 3.3 remain comparable due to the large uncertainties.

To quantitatively investigate the evolution, we derive the ${\rm SB_{Ly\alpha}}$ intrinsic profiles amplitudes at $r=200$ ckpc (${\rm SB_{Ly\alpha,intr}}$) as a function of redshift.
We fit the relation with ${\rm SB_{Ly\alpha,intr}}\propto(1+z)^b$, where $b$ is a constant, weighting ${\rm SB_{Ly\alpha,intr}}$ with $\sqrt{N_{\rm LAE}}$ at each redshift.
We find that the best-fit value of $b$ is 3.1, which implies that the intrinsic ${\rm SB_{Ly\alpha}}$ profile amplitudes increase toward high redshifts roughly by $(1+z)^3$ at a given radius in units of ckpc.
This trend might correspond to increasing density of hydrogen gas toward high redshifts due to the evolution of the cosmic volume.
Nevertheless, it is still difficult to draw a conclusion due to large uncertainties.
We cannot rule out other or additional possibilities, such as higher Ly$\alpha$ escape fractions toward high redshifts \citep[e.g.,][]{Hayes+11,Konno+16}.
It is also necessary to investigate potential impact of the cosmic reionization on the neutral hydrogen density at $r\sim100-1000$ ckpc around LAEs.
Deeper observations in the future will help to elucidate the evolution and its physical interpretation.

\subsection{Physical Origins of Extended Ly$\alpha$ Emission} \label{subsec:physical_origin}

In this section, we compare the observational results obtained in Section \ref{sec:im} with theoretical models taken from the literature to investigate the mechanism of extended Ly$\alpha$ emission production.
We investigate extended Ly$\alpha$ emission focusing on 1) where Ly$\alpha$ photons originate; 2) which processes produce Ly$\alpha$ photons; and 3) how these photons transfer in the surrounding materials.

1) First, we distinguish Ly$\alpha$ emission according to where it originates from:
\begin{enumerate}
\item{The ISM of the targeted galaxy ({\it central galaxy}).}
\item{The {\it CGM} surrounding the central galaxy.}
\item{{\it Satellite galaxies}.}
\item{{\it Other halos}, which refer to halos distinct from that hosting the central galaxy.}
\end{enumerate}

2) We consider that Ly$\alpha$ photons are produced in the processes of {\it recombination} and/or {\it collisional excitation} ({\it cooling radiation}), as stated in the Introduction.

3) Lastly, Ly$\alpha$ photons subsequently transfer through surrounding hydrogen gas while being resonantly scattered, which affects the observed SB profiles.
Hence, we should be conscious of whether models takes into account the scattering process or not (see \citealt{Byrohl+20} for the impact of scattering on ${\rm SB_{Ly\alpha}}$ profiles).

\begin{deluxetable*}{lcccccc}
\tablecaption{Summary of the theoretical studies used for comparison.
\label{tab:compare_model}}
\tablehead{\colhead{Reference} & \colhead{$z_{\rm model}$} & \colhead{$z_{\rm plot}$} & \colhead{Origin} & \colhead{Process} & \colhead{Scattering} & \colhead{$M_{\rm halo}\ (M_\odot)$}}
\colnumbers
\startdata
\citet{Mas-Ribas+17c}       & 5.7, 6.6   & 2.2, 3.3, 5.7, 6.6 & CGM/Sat.         & Rec.       & n & $10^{11\ *}$ \\
\citet{Kakiichi-Dijkstra18} & $2-3$      & 2.2                & Cen.             & Rec.       & y & $\sim10^{12}$ \\
\cite{Byrohl+20}            & $2-5$      & 3.3                & Cen./Sat./Other. & Rec./Cool. & y & $\sim10^{11\ \#}$ \\
\citet{Dijkstra-Kramer12}   & $\sim2.65$ & 3.3                & Cen.             & Rec.       & y & --- \\
\citet{Lake+15}             & 3.1        & 3.3                & Cen./Other.      & Rec./Cool. & y & $10^{11.5}$ \\
\citet{Zheng+11}            & 5.7        & 5.7                & Cen./Other.      & Rec.       & y & $10^{11.2}$ \\
\enddata
\tablecomments{Columns: (1) Reference. (2) Redshift assumed in the model. (3) Redshift where we plot the model in Figure \ref{fig:sb_lya_vsmodel_all}. (4) Origins of Ly$\alpha$ photons (where Ly$\alpha$ photons are produced): the central galaxy (Cen.), CGM, satellite galaxies (Sat.), and other halos (Other.) (5) Physical processes of Ly$\alpha$ emission production (how Ly$\alpha$ emission is produced): recombination (Rec.), and cooling radiation (Cool.). (6) Whether the model takes into account Ly$\alpha$ RT (i.e., resonant scattering) or not. (7) Assumed halo mass in units of $M_\odot$.\\
$^*$ We fix $M_{\rm halo}$ to $10^{11}\ M_\odot$, while $M_{\rm halo}$ is a free parameter in the model. \\
$^\#$ This value was converted from the assumed stellar mass $M_\star$ of $10^{8.5}-10^{9.5}\ M_\odot$ with the $M_\star/M_{\rm halo}$ ratio obtained in \citet{Behroozi+19}.}
\end{deluxetable*}

\begin{figure*}
\epsscale{1.1}
\plotone{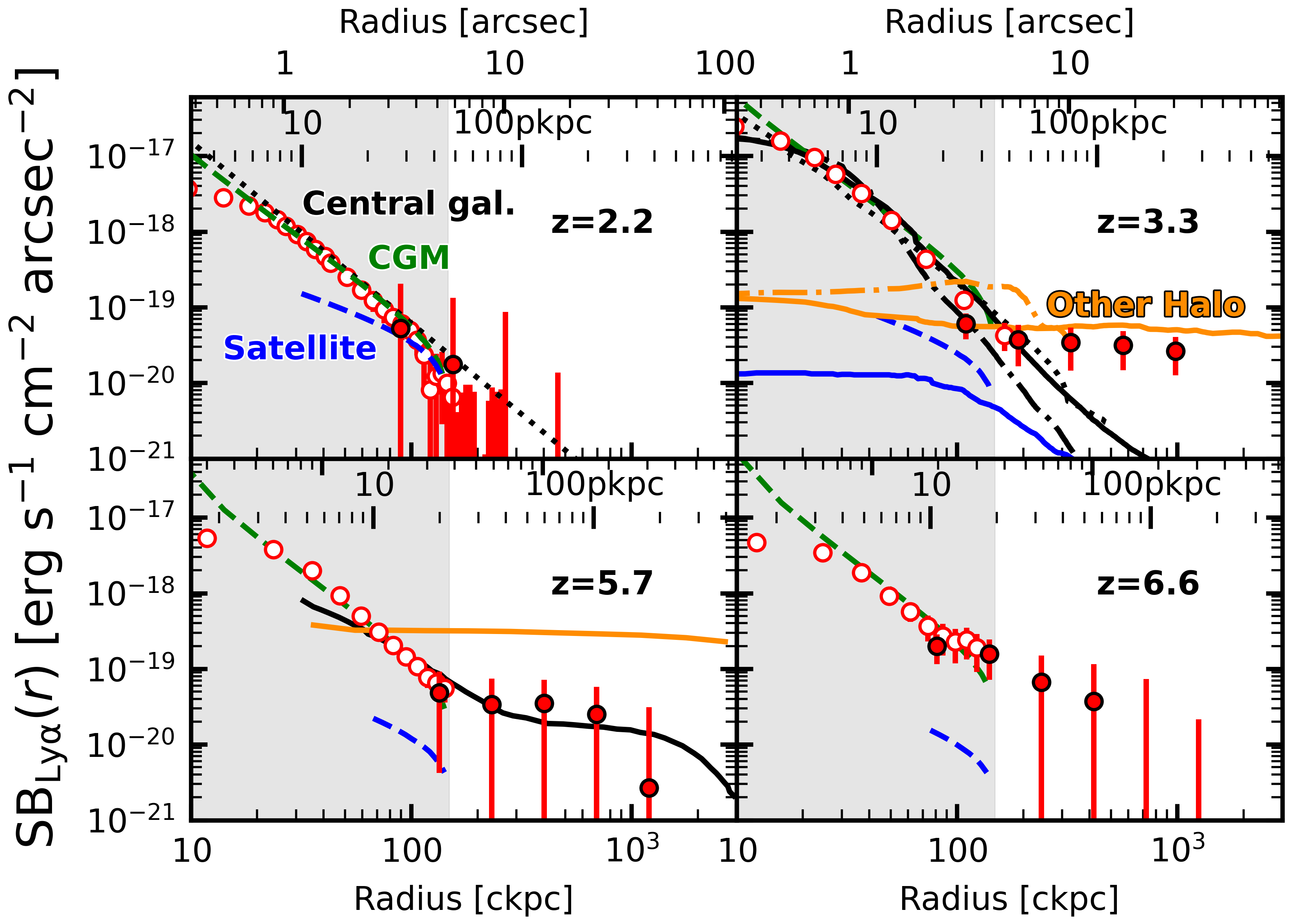}
\caption{Comparison of the ${\rm SB_{Ly\alpha}}$ radial profiles between the observational and theoretical studies at $z=2.2$ (top left), 3.3 (top right), 5.7 (bottom left), and 6.6 (bottom right). The filled red circles depict the observational results taken from our {\it all} sample ($z=2.2-6.6$), while the open red circles show the results from previous observational work (\citealt{Momose+16} at $z=2.2$, \citealt{Wisotzki+18} at $z=3.3$, and \citealt{Momose+14} at $z=5.7$ and 6.6). The ${\rm SB_{Ly\alpha}}$ amplitude of \citet{Wisotzki+18} is normalized such that $L_{\rm Lya}$ in $r\leq1\arcsec$ matches to that of our {\it all} sample (see Section \ref{subsec:compare_obs}). We represent the theoretical work with the lines, whose colors indicate the Ly$\alpha$ emission originations: the central galaxy (black), CGM (green), satellite galaxies (blue), and other halos (orange). The models are taken from \citet[green and blue dashed lines at $z=2.2$, 3.3, 5.7, 6.6]{Mas-Ribas+17c}, \citet[black dotted line at $z=2.2$]{Kakiichi-Dijkstra18}, \citet[black, blue, and orange solid lines at $z=3.3$]{Byrohl+20}, \citet[black and orange dash-dotted lines at $z=3.3$]{Lake+15}, \citet[black dotted line at $z=3.3$]{Dijkstra-Kramer12}, and \citet[black and orange solid lines at $z=5.7$]{Zheng+11}. See Table \ref{tab:compare_model} for details of these models. We normalize the ${\rm SB_{Ly\alpha}}$ profiles from the model predictions such that the central galaxy ($r\leq1\arcsec$) has the $L_{\rm Ly\alpha}$ value similar to that observed at each redshift. The gray shades illustrate the regions inside $R_{\rm vir}$ of a DMH with $M_{\rm halo}=10^{11}\ M_\odot$.}
\label{fig:sb_lya_vsmodel_all}
\end{figure*}

In Figure \ref{fig:sb_lya_vsmodel_all}, we compare the ${\rm SB_{Ly\alpha}}$ profiles that are observed (Section \ref{subsec:sb_lya}) and those predicted by theoretical work.
The observational results are taken from \citet[$L_{\rm Ly\alpha}=10^{42.6}$ erg s$^{-1}$ subsample at $z=2.2$]{Momose+16}, \citet[$L_{\rm Ly\alpha}>10^{42}$ erg s$^{-1}$ subsample at $z=3-4$]{Wisotzki+18}, \citet[$z=5.7$ and 6.6 LAEs]{Momose+14}, and our {\it all} samples ($z=2.2-6.6$).
Predicted profiles are taken from \citet{Zheng+11}, \citet{Dijkstra-Kramer12}, \citet{Lake+15}, \citet{Mas-Ribas+17c}, \citet{Kakiichi-Dijkstra18}, and \citet{Byrohl+20}, which are summarized in Table \ref{tab:compare_model}.
We plot these observational and theoretical results at the nearest redshifts among $z=2.2$, 3.3, 5.7, and 6.6, except that the model of \citet{Mas-Ribas+17c} is presented at all the redshifts, with the correction for the cosmological dimming effect.
We normalize all the profiles such that they match at $r=1\arcsec$ in amplitude for precise comparison under the same Ly$\alpha$ luminosity of the central galaxy (the models of \citealt{Zheng+11}, \citealt{Lake+15} and \citealt{Byrohl+20} are normalized in each total profile summing up different origins).
We compare these observational and theoretical results in the context of the Ly$\alpha$ photon origins in the following subsections.

\subsubsection{Central Galaxy} \label{subsubsec:central}

To discuss the contribution from the {\it central galaxy}, we compare the models of \citet{Zheng+11}, \citet{Dijkstra-Kramer12}, \citet{Lake+15}, \citet{Kakiichi-Dijkstra18}, and \citet{Byrohl+20}.
They applied Ly$\alpha$ radiative transfer (RT) modeling to hydrodynamic cosmological galaxy formation simulations at $z=5.7$, 3.1, $2-3$, and $2-5$, respectively, to investigate Ly$\alpha$ photons produced by SF in the central galaxy and resonantly scattered into the CGM.
We take the `one-halo' term model from \citet{Zheng+11}, where Ly$\alpha$ photons are scattered not only in the CGM but also the IGM.
The model of \citet{Lake+15} takes into account the contribution from cooling radiation in addition to that from SF.
The model of \citet{Kakiichi-Dijkstra18} is taken in an approximated form of their Equation (28).
The model of \citet{Byrohl+20} considers recombination caused by ionizing photons from SF and UVB as well as cooling via collisional de-excitation.

The black lines in Figure \ref{fig:sb_lya_vsmodel_all} represent the contribution from the central galaxy: \citet[dotted, $z=2.2$]{Kakiichi-Dijkstra18}, \citet[solid, $z=3.3$]{Byrohl+20}, \citet[dash-dotted, $z=3.3$]{Lake+15}, \citet[dotted, $z=3.3$]{Dijkstra-Kramer12}, and \citet[dotted, $z=5.7$]{Zheng+11}.
These models successfully reproduce the observed ${\rm SB_{Ly\alpha}}$ profiles inside the CGM ($r<R_{\rm vir}$) at $z=3.3$ and 5.7, implying resonant scattering as a major source powering Ly$\alpha$ emission.
This finding is also supported by previous studies in other aspects, such as halo properties (e.g., a halo luminosity-mass relation; \citealt{Kusakabe+19}) and kinematics (e.g., a correlation between the peak velocity shift and the width of a Ly$\alpha$ line, \citealt{Leclercq+20}; red peak dominated Ly$\alpha$ spectra, \citealt{Chen+21}).

The one-halo term of \citet{Zheng+11} reproduces the observed ${\rm SB_{Ly\alpha}}$ profile also from $r\sim R_{\rm vir}$ up to $\sim1$ cMpc at $z=5.7$.
They considered resonant scattering in the IGM in addition to the CGM.
This effect leads to a plateau-like feature in a ${\rm SB_{Ly\alpha}}$ profile at $r\sim0.3-1$ cMpc (see also \citealt{Jeeson-Daniel+12}), which is similar to those at $z=3.3$ and 5.7.
Therefore, we may interpret that the observed extended Ly$\alpha$ emission outside the CGM is produced by resonant scattering in the IGM, although the uncertainties are large.
At $z=3.3$, however, the models adopted here (\citealt{Lake+15}, \citealt{Dijkstra-Kramer12}, and \citealt{Byrohl+20}) produce values lying far below the observed ${\rm SB_{Ly\alpha}}$ profile beyond $300$ ckpc and do not reproduce a plateau-like shape, unlike the one-halo term of \citet{Zheng+11}.
One possibility is that the contribution from the central galaxy decreases from $z=5.7$ to 3.3, but there is no evidence to confirm this.
Alternatively, we suppose that this discrepancy can be attributed to the different assumptions and incorporated physics in the models, such as treatment of stellar radiation, dusts, and scattering, especially beyond $R_{\rm vir}$.
Nevertheless, since the one-halo term of \citet{Zheng+11} reproduces the observed ${\rm SB_{Ly\alpha}}$ profile at $z=5.7$, we cannot rule out the possibility that scattered Ly$\alpha$ photons originating from the central galaxy contribute to extended Ly$\alpha$ emission beyond $R_{\rm vir}$.
We need additional inputs on the neutral hydrogen gas distribution outside $R_{\rm vir}$ to further determine the contribution of resonant scattering.

The assumptions on $M_{\rm halo}$ values are unlikely to affect our discussion here, because the models above use roughly similar $M_{\rm halo}$ values: $10^{11.2}$ and $10^{11.5}\ M_\odot$ for \citet{Zheng+11} and \citet{Lake+15}, respectively.
\citet{Byrohl+20} assume a stellar mass $M_\star$ range of $10^{8.5}-10^{9.5}\ M_\odot$ at $z=3$, which corresponds to $M_{\rm halo}\sim10^{11}\ M_\odot$ given the $M_\star/M_{\rm halo}$ ratio obtained in \citet[see also \citealt{Kusakabe+18}]{Behroozi+19}.
These values are similar to those obtained in the previous observations \citep[e.g.,][]{Ouchi+10,Kusakabe+18}.

\subsubsection{CGM} \label{subsubsec:cgm}

We next use the model from \citet{Mas-Ribas+17c} to investigate the Ly$\alpha$ emission produced in the {\it CGM}.
They constructed an analytical model of fluorescent emission in the CGM caused by ionizing radiation from SF in the central galaxy at $z=5.7$ and 6.6 (we apply this model also to $z=3.3$ and 2.2; see also \citealt{Mas-Ribas-Dijkstra16}).
The model of \citet{Mas-Ribas+17c} includes three free parameters: 1) the CGM structure, 2) SF rate (SFR), and 3) radius $R_{\rm max}$.
1) We adopt the simplified clumpy outflow model of \citet{Steidel+10} as the CGM structure (the choice here has only a small impact on ${\rm SB_{Ly\alpha}}$ profiles; see Figure 2 of \citealt{Mas-Ribas-Dijkstra16}).
2) We normalize the model with ${\rm SFR}=1$, 10, 10, and 20 $M_\odot$ yr$^{-1}$ at $z=2.2$, 3.3, 5.7, 6.6, respectively.
3) ${\rm SB_{Ly\alpha}}(b)$ is derived as the integration of Ly$\alpha$ emissivity at radius $r$ over $b\leq r\leq R_{\rm max}$, where $b$ denotes the impact parameter (see Equations 4 and 2 of \citealt{Mas-Ribas-Dijkstra16} and \citealt{Mas-Ribas+17c}, respectively).
Here $R_{\rm max}$ represents the extent to which Ly$\alpha$ emission contributes to ${\rm SB_{Ly\alpha}}$.
We assume that $R_{\rm max}$ is equivalent to $R_{\rm vir}$ for a DMH of $M_{\rm halo}=10^{11}\ M_\odot$ ($=46$, 34, 22, and 20 pkpc at $z=2.2$, 3.3, 5.7, and 6.6, respectively).
Their model ignores the effect of resonant scattering.

The green dashed lines in Figure \ref{fig:sb_lya_vsmodel_all} represent the model of \citet{Mas-Ribas+17c}.
The ${\rm SB_{Ly\alpha}}$ profiles predicted by their model are in good agreement with those observed inside $R_{\rm vir}$ at all the redshifts.
However, because ${\rm SB_{Ly\alpha}}$ profiles drops sharply at $r=R_{\rm max}$ according to this model, the emission originating from the CGM cannot contribute to extended emission beyond $R_{\rm vir}$.
\citet{Mas-Ribas-Dijkstra16} and \citet{Mas-Ribas+17c} arbitrarily adopted much larger $R_{\rm max}$ values ($>100$ pkpc) to reproduce the profiles obtained in \citet{Momose+14}, bust such large $R_{\rm max}$ values correspond to $M_{\rm halo}>10^{12}\ M_\odot$, which is much larger than those observed \citep[e.g.,][]{Ouchi+10,Kusakabe+18}.
If $R_{\rm max}$ is larger than $R_{\rm vir}$, materials should exist outside the CGM ($\gtrsim3$ times larger scales than $R_{\rm vir}$) and produce fluorescent emission contributing to extended Ly$\alpha$ emission.
In either case, the observed ${\rm SB_{Ly\alpha}}$ profiles beyond 300 cpkc at $z=3.3$ cannot be reproduced even with larger values of $R_{\rm max}$.

Overall, fluorescence in the CGM can power Ly$\alpha$ emission inside $R_{\rm vir}$ according to the model of \citet{Mas-Ribas+17c}, while it plays only a marginal role beyond $R_{\rm vir}$.
This behavior was also suggested by MUSE UDF data \citep{Gallego+18,Bacon+21}.
We note that the model of \citet{Mas-Ribas+17c} ignores the scattering effect, which leads to a sharp drop of ${\rm SB_{Ly\alpha}}$ at $r=R_{\rm max}$.
Hence, it is necessary to incorporate resonant scattering to extend Ly$\alpha$ emission when we rely on the CGM fluorescence scenario.

The contribution from cooling radiation in the CGM remains unclear.
\citet{Byrohl+20} argued that cooling radiation dominates $\sim30$ \% of the total Ly$\alpha$ emission at $r\gtrsim20$ pkpc.
On the other hand, \citet{Rosdahl-Blaizot12} found that Ly$\alpha$ emission from cooling radiation is centrally ($r<10$ pkpc) concentrated for a DMH with $M_{\rm halo}=10^{11}\ M_\odot$.
We need additional models at larger scales to further discuss whether cooling radiation contributes to extended Ly$\alpha$ emission or not.

\subsubsection{Satellite Galaxies} \label{subsubsec:satellite}

We adopt another model from \citet{Mas-Ribas+17c}, which predicts the contribution from SF in {\it satellite galaxies} (see also \citealt{Mas-Ribas+17a}).
There are three free parameters in their model: 1) a clustering description, 2) Ly$\alpha$ escape fraction ($f_{\rm Lya}^{\rm esc}$), and 3) $R_{\rm max}$.
1) We assume that clustering follows a power-law two-point cross-correlation function $\xi(r)$ of $\xi(r)=(r/r_0)^{-\alpha}$ with the scale length $r_0=4$ cMpc and index $\alpha=1.8$ \citep[e.g.,][]{Ouchi+10,Harikane+16,Bielby+17}.
2) We fix $f_{\rm Lya}^{\rm esc}$ to 0.4, while ${\rm SB_{Ly\alpha}}$ linearly depends on $f_{\rm Lya}^{\rm esc}$.
3) Satellite galaxies are assumed to exist from $r=10$ pkpc to $R_{\rm max}$, which we assume equal to $R_{\rm vir}$ in the same way as for the fluorescence model (Section \ref{subsubsec:cgm}).
We additionally take the `outer halo' model from \citet{Byrohl+20}.

The blue dashed and solid lines represent the models of \citet{Mas-Ribas+17c} and \citet{Byrohl+20}, respectively.
We find that the contribution from satellite galaxies are negligible compared to the other contributions, except at $r\sim100$ ckpc at $z=2.2$.
While we choose a power-low correlation function to describe the clustering, other choices, such as the Navarro-Frenk-White (NFW) profile \citep{Navarro+97}, reduce the ${\rm SB_{Ly\alpha}}$ values at $r>40$ ckpc (see the left panel of Figure 2 of \citealt{Mas-Ribas+17a}).
When a power-law correlation function is assumed, larger $r_0$ values increase the overall profiles.
However, unrealistically large values of $r_0$ and $R_{\rm max}$ are necessary to reproduce extended Ly$\alpha$ emission beyond $R_{\rm vir}$ with satellite galaxies alone.
Additionally, the model overpredicts the observed ${\rm SB_{cont,\nu}}$ values when the model is tuned to reproduce the observed ${\rm SB_{Ly\alpha}}$ profiles \citep{Mas-Ribas+17a}.

For these reasons, we conclude that satellite galaxies are unlikely to contribute to extended Ly$\alpha$ emission beyond $R_{\rm vir}$.
This conclusion is supported by the fact that emission is more extended in Ly$\alpha$ than in UV continuum (Section \ref{subsec:sb_lya}; see also \citealt{Momose+14,Momose+16,Wu+20}), because the ${\rm SB_{cont,\nu}}$ profiles should be extended similarly as the ${\rm SB_{Ly\alpha}}$ profiles if satellite galaxies contribute to ${\rm SB_{Ly\alpha}}$.

\subsubsection{Other Halos} \label{subsubsec:otherhalo}

Lastly, we compare the models for {\it other halos} taken from \citet{Zheng+11}, \citet{Lake+15}, and \citet{Byrohl+20}.
From \citet{Lake+15} we specifically adopt the model in which Ly$\alpha$ emission originates from `knots,' the regions with high Ly$\alpha$ emissivity around the central galaxy.
The `two-halo' term model is taken from \citet{Zheng+11}.

We show these models with the orange lines in Figure \ref{fig:sb_lya_vsmodel_all}: \citet[solid, $z=3.3$]{Byrohl+20}, \citet[dash-dotted, $z=3.3$]{Lake+15}, and \citet[solid, $z=5.7$]{Zheng+11}.
At $z=3.3$, the contribution from other halos predicted by \citet{Byrohl+20} agrees with the observed ${\rm SB_{Ly\alpha}}$ profiles within the $1\sigma$ uncertainties.
Although the knots model of \citet{Lake+15} is limited to $r\lesssim300$ cpkc, it roughly reproduces the observed profiles at $r\sim(200-300)$ ckpc.
On the other hand, the two-halo term of \citet{Zheng+11} at $z=5.7$ significantly overestimates the ${\rm SB_{Ly\alpha}}$ values beyond 100 ckpc.
The amplitudes of the models of \citet{Zheng+11} and \citet{Byrohl+20} differ by $\sim1$ dex, similarly as we found in Section \ref{subsubsec:central}.

\citet{Kakuma+21} argued that the difference is caused because they masked out bright objects.
However, this interpretation is not necessarily appropriate, since the profile of \citet{Byrohl+20} has an amplitude similar to (or rather slightly higher than) profiles observed at $z=3.3$.
The two-halo term of \citet{Zheng+11} also overpredicts the ${\rm SB_{Ly\alpha}}$ values of \cite{Momose+14}.
Nevertheless, we cannot rule out the possibility that other halos contribute to extended Ly$\alpha$ emission beyond $R_{\rm vir}$, since the model of \citet{Byrohl+20} agrees with the observed ${\rm SB_{Ly\alpha}}$ profiles.
This suggestion is consistent with \citet{Bacon+21}, who identified very extended ($>300$ arcsec$^2$ or $>2\times10^4$ pkpc$^2$) Ly$\alpha$ emission at $z\sim3$ using MUSE data; they found that 70 \% of the total Ly$\alpha$ luminosity originates from filamentary structures beyond the CGM.
They argued that the extended Ly$\alpha$ emission can be reproduced by a population of extremely faint ($<10^{40}$ erg s$^{-1}$) galaxies under certain conditions, which correspond to other halos considered in this subsection.

We note that the ${\rm SB_{Ly\alpha}}$ profile of our {\it all} sample at $z=5.7$ increases when we use only LAEs in the UD-COSMOS field, resulting in a smaller amplitude gap between the two-halo term of \citet{Zheng+11} and our profile.
However, our profile based solely on the UD-COSMOS field disagrees with those from the previous observational studies.
We thus suppose that the averaged profile is more appropriate for comparison against the models.

\subsubsection{Overall Interpretation} \label{subsubsec:summary}

In summary, ${\rm SB_{Ly\alpha}}$ profiles inside the CGM ($<R_{\rm vir}$) are possibly explained either by scattered Ly$\alpha$ emission originating from the central galaxy and/or fluorescent emission in the CGM.
Meanwhile, extended Ly$\alpha$ emission beyond $R_{\rm vir}$ is possibly powered either by resonant scattering at large scales and/or contributed from other halos.
Fluorescence in the CGM and satellite galaxies are not sufficient to reproduce the observed ${\rm SB_{Ly\alpha}}$ profiles beyond $R_{\rm vir}$.

We emphasize that the processes and origins of Ly$\alpha$ emission may differ among LAEs.
They may also vary according to radius and redshift even when we focus on averaged profiles around different LAEs.
However, our systematic investigation of extended Ly$\alpha$ emission at $z=2-7$ is advantageous for a comprehensive understanding of the processes and origins of extended Ly$\alpha$ emission.
More simulations focusing on large scales will help to further distinguish the processes and origins.

\section{Summary} \label{sec:summary}

In this paper, we investigated very extended Ly$\alpha$ emission around LAEs at $z=2.2-6.6$ by applying the intensity mapping technique to the Subaru/HSC-SSP and CHORUS data.
Our major findings are summarized below:

\begin{enumerate}
\item{We calculated cross-correlation functions between 1781 LAEs at $z=2.2$, 3.3, 5.7 and 6.6 with Ly$\alpha$ emission traced by the NB387, NB527, NB816, and NB921 images. A total of $\sim1-2$ billion pixels were used to derive the correlation function for each redshift. The deep ($m_{\rm NB,5\sigma}\sim26$ mag) and wide ($\sim4$ deg$^2$ over the UD-COSMOS and UD-SXDS fields) images of the HSC enabled us to detect very diffuse Ly$\alpha$ emission. We utilized foreground objects (NonLAEs) to carefully estimate the systematics, including the sky background and PSF. Subtracting these systematics, we identified the Ly$\alpha$ emission of $\sim10^{-20}-10^{-19}$ erg s$^{-1}$ cm$^{-2}$ arcsec$^{-2}$ with ${\rm S/N}=4.1$ around the $z=3.3$ LAEs. Ly$\alpha$ emission extends beyond the radial scale of the $R_{\rm vir}$ of a DMH with $10^{11}\ M_\odot$ ($\sim100$ ckpc), and up to $\sim1$ cMpc. We also tentatively detected Ly$\alpha$ emission beyond the $R_{\rm vir}$ scales at $z=5.7$, and 6.6 with ${\rm S/N}\sim2$. Extended Ly$\alpha$ emission was also tentatively detected around the $z=2.2$ LAEs when faint LAEs were excluded from the sample.}

\item{We confirmed that the Ly$\alpha$ surface brightness (${\rm SB_{Ly\alpha}}$) radial profiles around our LAEs agree well with those obtained in the previous studies, when the LAEs have similar Ly$\alpha$ luminosity values.}

\item{We compared the observed ${\rm SB_{Ly\alpha}}$ profiles across $z=2.2-6.6$, finding no significant difference among the redshifts beyond the uncertainties. Meanwhile, there is a potential increasing trend toward high redshifts in the {\it intrinsic} ${\rm SB_{Ly\alpha}}$ profiles, which are corrected for the cosmological dimming effect. The increasing trend roughly follows $(1+z)^3$, which might be explained by the increasing density of the neutral hydrogen gas due to the evolution of the cosmic volume.}

\item{We compared the ${\rm SB_{Ly\alpha}}$ profiles obtained from the observational and theoretical studies. We found that the observed ${\rm SB_{Ly\alpha}}$ profiles inside the CGM can be reproduced by the models in which Ly$\alpha$ photons originating from the central galaxy subsequently transfer into the CGM via resonant scattering, or in which Ly$\alpha$ emission is produced in the CGM via fluorescence due to ionizing photons. Extended Ly$\alpha$ emission beyond $R_{\rm vir}$ may be reproduced by resonant scattering at large scales, and/or emission originating from clustered halos around the targeted galaxy. The CGM and satellite galaxies are unlikely to contribute to extended Ly$\alpha$ emission beyond $R_{\rm vir}$.}

\end{enumerate}

This work, in conjunction with the previous observational studies, might suggest that very extended diffuse Ly$\alpha$ emission beyond $R_{\rm vir}$ ubiquitously exist around LAEs at $z\sim2-7$, not only around massive galaxies.
Deeper images obtained by larger-area surveys in the future should enable further investigation of very extended Ly$\alpha$ emission at more diffuse levels.
Applying the intensity mapping technique to the emission of multiple lines, such as H$\alpha$ and [{\sc Oiii}], will help to further distinguish physical processes and origins of extended Ly$\alpha$ emission, because they trace different components (see Figures 6 and 12 of \citealt{Mas-Ribas+17a} and \citealt{Fujimoto+19}).
These emission lines will be observed with next-generation facilities, such as the James-Webb Space Telescope (JWST), the Nancy Grace Roman Space Telescope (NGRST), and Spectro-Photometer for the History of the Universe, Epoch of Reionization and Ices Explorer (SPHEREx).

\section*{Acknowledgements}

We are grateful to Chris Byrohl for helpful comments and discussions especially on simulations and physical interpretations.

The HSC collaboration includes the astronomical communities of Japan and Taiwan, and Princeton University. The HSC instrumentation and software were developed by the National Astronomical Observatory of Japan (NAOJ), the Kavli Institute for the Physics and Mathematics of the Universe (Kavli IPMU), the University of Tokyo, the High Energy Accelerator Research Organization (KEK), the Academia Sinica Institute for Astronomy and Astrophysics in Taiwan (ASIAA), and Princeton University. Funding was contributed by the FIRST program from the Japanese Cabinet Office, the Ministry of Education, Culture, Sports, Science and Technology (MEXT), the Japan Society for the Promotion of Science (JSPS), Japan Science and Technology Agency (JST), the Toray Science Foundation, NAOJ, Kavli IPMU, KEK, ASIAA, and Princeton University.

The Pan-STARRS1 Surveys (PS1) and the PS1 public science archive have been made possible through contributions by the Institute for Astronomy, the University of Hawaii, the Pan-STARRS Project Office, the Max Planck Society and its participating institutes, the Max Planck Institute for Astronomy, Heidelberg, and the Max Planck Institute for Extraterrestrial Physics, Garching, The Johns Hopkins University, Durham University, the University of Edinburgh, the Queen’s University Belfast, the Harvard-Smithsonian Center for Astrophysics, the Las Cumbres Observatory Global Telescope Network Incorporated, the National Central University of Taiwan, the Space Telescope Science Institute, the National Aeronautics and Space Administration under grant No. NNX08AR22G issued through the Planetary Science Division of the NASA Science Mission Directorate, the National Science Foundation grant No. AST-1238877, the University of Maryland, Eotvos Lorand University (ELTE), the Los Alamos National Laboratory, and the Gordon and Betty Moore Foundation.

This paper makes use of software developed for the Large Synoptic Survey Telescope. We thank the LSST Project for making their code available as free software at \url{http://dm.lsst.org}.

This paper is based on data collected at the Subaru Telescope and retrieved from the HSC data archive system, which is operated by the Subaru Telescope and Astronomy Data Center (ADC) at National Astronomical Observatory of Japan. Data analysis was in part carried out with the cooperation of Center for Computational Astrophysics (CfCA), National Astronomical Observatory of Japan. The Subaru Telescope is honored and grateful for the opportunity of observing the Universe from Maunakea, which has the cultural, historical and natural significance in Hawaii. 

The NB387 filter was supported by KAKENHI (23244022) Grant-in-Aid for Scientific Research (A) through the JSPS.
The NB527 filter was supported by KAKENHI (24244018) Grant-in-Aid for Scientific Research (A) through the JSPS.
The NB816 filter was supported by Ehime University.
The NB921 filter was supported by KAKENHI (23244025) Grant-in-Aid for Scientific Research (A) through the JSPS.

This work is supported by the World Premier International Research Center Initiative (WPI Initiative), MEXT, Japan, as well as KAKENHI Grant-in-Aid for Scientific Research (A) (20H00180, and 21H04467) through the JSPS.
This work was supported by JSPS KAKENHI Grant Numbers 17H01114, 19J01222, 20J11993, and 21K13953.

This work was supported by the joint research program of the Institute for Cosmic Ray Research (ICRR), University of Tokyo.


\bibliographystyle{aasjournal}
\bibliography{cite.bib}

\begin{thebibliography}{}
\expandafter\ifx\csname natexlab\endcsname\relax\def\natexlab#1{#1}\fi
\providecommand{\url}[1]{\href{#1}{#1}}
\providecommand{\dodoi}[1]{doi:~\href{http://doi.org/#1}{\nolinkurl{#1}}}
\providecommand{\doeprint}[1]{\href{http://ascl.net/#1}{\nolinkurl{http://ascl.net/#1}}}
\providecommand{\doarXiv}[1]{\href{https://arxiv.org/abs/#1}{\nolinkurl{https://arxiv.org/abs/#1}}}

\bibitem[{{Aihara} {et~al.}(2019){Aihara}, {AlSayyad}, {Ando}, {Armstrong},
  {Bosch}, {Egami}, {Furusawa}, {Furusawa}, {Goulding}, {Harikane}, {Hikage},
  {Ho}, {Hsieh}, {Huang}, {Ikeda}, {Imanishi}, {Ito}, {Iwata}, {Jaelani},
  {Kakuma}, {Kawana}, {Kikuta}, {Kobayashi}, {Koike}, {Komiyama}, {Li},
  {Liang}, {Lin}, {Luo}, {Lupton}, {Lust}, {MacArthur}, {Matsuoka}, {Mineo},
  {Miyatake}, {Miyazaki}, {More}, {Murata}, {Namiki}, {Nishizawa}, {Oguri},
  {Okabe}, {Okamoto}, {Okura}, {Ono}, {Onodera}, {Onoue}, {Osato}, {Ouchi},
  {Shibuya}, {Strauss}, {Sugiyama}, {Suto}, {Takada}, {Takagi}, {Takata},
  {Takita}, {Tanaka}, {Terai}, {Toba}, {Uchiyama}, {Utsumi}, {Wang}, {Wang}, \&
  {Yamada}}]{Aihara+19}
{Aihara}, H., {AlSayyad}, Y., {Ando}, M., {et~al.} 2019, \pasj, 71, 114,
  \dodoi{10.1093/pasj/psz103}

\bibitem[{{Arrigoni Battaia} {et~al.}(2019){Arrigoni Battaia}, {Hennawi},
  {Prochaska}, {O{\~n}orbe}, {Farina}, {Cantalupo}, \&
  {Lusso}}]{Arrigoni-Battaia+19}
{Arrigoni Battaia}, F., {Hennawi}, J.~F., {Prochaska}, J.~X., {et~al.} 2019,
  \mnras, 482, 3162, \dodoi{10.1093/mnras/sty2827}

\bibitem[{{Bacon} {et~al.}(2021){Bacon}, {Mary}, {Garel}, {Blaizot}, {Maseda},
  {Schaye}, {Wisotzki}, {Conseil}, {Brinchmann}, {Leclercq}, {Abril-Melgarejo},
  {Boogaard}, {Bouch{\'e}}, {Contini}, {Feltre}, {Guiderdoni}, {Herenz},
  {Kollatschny}, {Kusakabe}, {Matthee}, {Michel-Dansac}, {Nanayakkara},
  {Richard}, {Roth}, {Schmidt}, {Steinmetz}, {Tresse}, {Urrutia}, {Verhamme},
  {Weilbacher}, {Zabl}, \& {Zoutendijk}}]{Bacon+21}
{Bacon}, R., {Mary}, D., {Garel}, T., {et~al.} 2021, \aap, 647, A107,
  \dodoi{10.1051/0004-6361/202039887}

\bibitem[{{Behroozi} {et~al.}(2019){Behroozi}, {Wechsler}, {Hearin}, \&
  {Conroy}}]{Behroozi+19}
{Behroozi}, P., {Wechsler}, R.~H., {Hearin}, A.~P., \& {Conroy}, C. 2019,
  \mnras, 488, 3143, \dodoi{10.1093/mnras/stz1182}

\bibitem[{{Bertin} \& {Arnouts}(1996)}]{Bertin-Arnouts96}
{Bertin}, E., \& {Arnouts}, S. 1996, \aaps, 117, 393,
  \dodoi{10.1051/aas:1996164}

\bibitem[{{Bielby} {et~al.}(2017){Bielby}, {Shanks}, {Crighton}, {Bornancini},
  {Infante}, {Lambas}, {Minniti}, {Morris}, \& {Tummuangpak}}]{Bielby+17}
{Bielby}, R.~M., {Shanks}, T., {Crighton}, N.~H.~M., {et~al.} 2017, \mnras,
  471, 2174, \dodoi{10.1093/mnras/stx1772}

\bibitem[{{Bond} {et~al.}(2010){Bond}, {Feldmeier}, {Matkovi{\'c}}, {Gronwall},
  {Ciardullo}, \& {Gawiser}}]{Bond+10}
{Bond}, N.~A., {Feldmeier}, J.~J., {Matkovi{\'c}}, A., {et~al.} 2010, \apjl,
  716, L200, \dodoi{10.1088/2041-8205/716/2/L200}

\bibitem[{{Borisova} {et~al.}(2016){Borisova}, {Cantalupo}, {Lilly}, {Marino},
  {Gallego}, {Bacon}, {Blaizot}, {Bouch{\'e}}, {Brinchmann}, {Carollo},
  {Caruana}, {Finley}, {Herenz}, {Richard}, {Schaye}, {Straka}, {Turner},
  {Urrutia}, {Verhamme}, \& {Wisotzki}}]{Borisova+16}
{Borisova}, E., {Cantalupo}, S., {Lilly}, S.~J., {et~al.} 2016, \apj, 831, 39,
  \dodoi{10.3847/0004-637X/831/1/39}

\bibitem[{{Bosch} {et~al.}(2018){Bosch}, {Armstrong}, {Bickerton}, {Furusawa},
  {Ikeda}, {Koike}, {Lupton}, {Mineo}, {Price}, {Takata}, {Tanaka}, {Yasuda},
  {AlSayyad}, {Becker}, {Coulton}, {Coupon}, {Garmilla}, {Huang}, {Krughoff},
  {Lang}, {Leauthaud}, {Lim}, {Lust}, {MacArthur}, {Mandelbaum}, {Miyatake},
  {Miyazaki}, {Murata}, {More}, {Okura}, {Owen}, {Swinbank}, {Strauss},
  {Yamada}, \& {Yamanoi}}]{Bosch+18}
{Bosch}, J., {Armstrong}, R., {Bickerton}, S., {et~al.} 2018, \pasj, 70, S5,
  \dodoi{10.1093/pasj/psx080}

\bibitem[{{Byrohl} {et~al.}(2020){Byrohl}, {Nelson}, {Behrens}, {Pillepich},
  {Hernquist}, {Marinacci}, \& {Vogelsberger}}]{Byrohl+20}
{Byrohl}, C., {Nelson}, D., {Behrens}, C., {et~al.} 2020, arXiv e-prints,
  arXiv:2009.07283.
\newblock \doarXiv{2009.07283}

\bibitem[{{Cantalupo} {et~al.}(2014){Cantalupo}, {Arrigoni-Battaia},
  {Prochaska}, {Hennawi}, \& {Madau}}]{Cantalupo+14}
{Cantalupo}, S., {Arrigoni-Battaia}, F., {Prochaska}, J.~X., {Hennawi}, J.~F.,
  \& {Madau}, P. 2014, \nat, 506, 63, \dodoi{10.1038/nature12898}

\bibitem[{{Cantalupo} {et~al.}(2005){Cantalupo}, {Porciani}, {Lilly}, \&
  {Miniati}}]{Cantalupo+05}
{Cantalupo}, S., {Porciani}, C., {Lilly}, S.~J., \& {Miniati}, F. 2005, \apj,
  628, 61, \dodoi{10.1086/430758}

\bibitem[{{Carilli}(2011)}]{Carilli11}
{Carilli}, C.~L. 2011, \apjl, 730, L30, \dodoi{10.1088/2041-8205/730/2/L30}

\bibitem[{{Chen} {et~al.}(2021){Chen}, {Steidel}, {Erb}, {Law}, {Trainor},
  {Reddy}, {Shapley}, {Pahl}, {Strom}, {Li}, \& {Rudie}}]{Chen+21}
{Chen}, Y., {Steidel}, C.~C., {Erb}, D.~K., {et~al.} 2021, arXiv e-prints,
  arXiv:2104.10173.
\newblock \doarXiv{2104.10173}

\bibitem[{{Claeyssens} {et~al.}(2019){Claeyssens}, {Richard}, {Blaizot},
  {Garel}, {Leclercq}, {Patr{\'\i}cio}, {Verhamme}, {Wisotzki}, {Bacon},
  {Carton}, {Cl{\'e}ment}, {Herenz}, {Marino}, {Muzahid}, {Saust}, \&
  {Schaye}}]{Claeyssens+19}
{Claeyssens}, A., {Richard}, J., {Blaizot}, J., {et~al.} 2019, \mnras, 489,
  5022, \dodoi{10.1093/mnras/stz2492}

\bibitem[{{Comaschi} \& {Ferrara}(2016{\natexlab{a}})}]{Comaschi-Ferrara16a}
{Comaschi}, P., \& {Ferrara}, A. 2016{\natexlab{a}}, \mnras, 455, 725,
  \dodoi{10.1093/mnras/stv2339}

\bibitem[{{Comaschi} \& {Ferrara}(2016{\natexlab{b}})}]{Comaschi-Ferrara+16b}
---. 2016{\natexlab{b}}, \mnras, 463, 3078, \dodoi{10.1093/mnras/stw2199}

\bibitem[{{Croft} {et~al.}(2018){Croft}, {Miralda-Escud{\'e}}, {Zheng},
  {Blomqvist}, \& {Pieri}}]{Croft+18}
{Croft}, R. A.~C., {Miralda-Escud{\'e}}, J., {Zheng}, Z., {Blomqvist}, M., \&
  {Pieri}, M. 2018, \mnras, 481, 1320, \dodoi{10.1093/mnras/sty2302}

\bibitem[{{Croft} {et~al.}(2016){Croft}, {Miralda-Escud{\'e}}, {Zheng},
  {Bolton}, {Dawson}, {Peterson}, {York}, {Eisenstein}, {Brinkmann},
  {Brownstein}, {Cen}, {Delubac}, {Font-Ribera}, {Hamilton}, {Lee}, {Myers},
  {Palanque-Delabrouille}, {P{\^a}ris}, {Petitjean}, {Pieri}, {Ross}, {Rossi},
  {Schlegel}, {Schneider}, {Slosar}, {Vazquez}, {Viel}, {Weinberg}, \&
  {Y{\`e}che}}]{Croft+16}
{Croft}, R. A.~C., {Miralda-Escud{\'e}}, J., {Zheng}, Z., {et~al.} 2016,
  \mnras, 457, 3541, \dodoi{10.1093/mnras/stw204}

\bibitem[{{Dawson} {et~al.}(2013){Dawson}, {Schlegel}, {Ahn}, {Anderson},
  {Aubourg}, {Bailey}, {Barkhouser}, {Bautista}, {Beifiori}, {Berlind},
  {Bhardwaj}, {Bizyaev}, {Blake}, {Blanton}, {Blomqvist}, {Bolton}, {Borde},
  {Bovy}, {Brandt}, {Brewington}, {Brinkmann}, {Brown}, {Brownstein}, {Bundy},
  {Busca}, {Carithers}, {Carnero}, {Carr}, {Chen}, {Comparat}, {Connolly},
  {Cope}, {Croft}, {Cuesta}, {da Costa}, {Davenport}, {Delubac}, {de Putter},
  {Dhital}, {Ealet}, {Ebelke}, {Eisenstein}, {Escoffier}, {Fan}, {Filiz Ak},
  {Finley}, {Font-Ribera}, {G{\'e}nova-Santos}, {Gunn}, {Guo}, {Haggard},
  {Hall}, {Hamilton}, {Harris}, {Harris}, {Ho}, {Hogg}, {Holder}, {Honscheid},
  {Huehnerhoff}, {Jordan}, {Jordan}, {Kauffmann}, {Kazin}, {Kirkby}, {Klaene},
  {Kneib}, {Le Goff}, {Lee}, {Long}, {Loomis}, {Lundgren}, {Lupton}, {Maia},
  {Makler}, {Malanushenko}, {Malanushenko}, {Mandelbaum}, {Manera}, {Maraston},
  {Margala}, {Masters}, {McBride}, {McDonald}, {McGreer}, {McMahon}, {Mena},
  {Miralda-Escud{\'e}}, {Montero-Dorta}, {Montesano}, {Muna}, {Myers},
  {Naugle}, {Nichol}, {Noterdaeme}, {Nuza}, {Olmstead}, {Oravetz}, {Oravetz},
  {Owen}, {Padmanabhan}, {Palanque-Delabrouille}, {Pan}, {Parejko},
  {P{\^a}ris}, {Percival}, {P{\'e}rez-Fournon}, {P{\'e}rez-R{\`a}fols},
  {Petitjean}, {Pfaffenberger}, {Pforr}, {Pieri}, {Prada}, {Price-Whelan},
  {Raddick}, {Rebolo}, {Rich}, {Richards}, {Rockosi}, {Roe}, {Ross}, {Ross},
  {Rossi}, {Rubi{\~n}o-Martin}, {Samushia}, {S{\'a}nchez}, {Sayres}, {Schmidt},
  {Schneider}, {Sc{\'o}ccola}, {Seo}, {Shelden}, {Sheldon}, {Shen}, {Shu},
  {Slosar}, {Smee}, {Snedden}, {Stauffer}, {Steele}, {Strauss}, {Streblyanska},
  {Suzuki}, {Swanson}, {Tal}, {Tanaka}, {Thomas}, {Tinker}, {Tojeiro},
  {Tremonti}, {Vargas Maga{\~n}a}, {Verde}, {Viel}, {Wake}, {Watson}, {Weaver},
  {Weinberg}, {Weiner}, {West}, {White}, {Wood-Vasey}, {Yeche}, {Zehavi},
  {Zhao}, \& {Zheng}}]{Dawson+13}
{Dawson}, K.~S., {Schlegel}, D.~J., {Ahn}, C.~P., {et~al.} 2013, \aj, 145, 10,
  \dodoi{10.1088/0004-6256/145/1/10}

\bibitem[{{Dijkstra} \& {Kramer}(2012)}]{Dijkstra-Kramer12}
{Dijkstra}, M., \& {Kramer}, R. 2012, \mnras, 424, 1672,
  \dodoi{10.1111/j.1365-2966.2012.21131.x}

\bibitem[{{Eisenstein} {et~al.}(2011){Eisenstein}, {Weinberg}, {Agol},
  {Aihara}, {Allende Prieto}, {Anderson}, {Arns}, {Aubourg}, {Bailey},
  {Balbinot}, {Barkhouser}, {Beers}, {Berlind}, {Bickerton}, {Bizyaev},
  {Blanton}, {Bochanski}, {Bolton}, {Bosman}, {Bovy}, {Brandt}, {Breslauer},
  {Brewington}, {Brinkmann}, {Brown}, {Brownstein}, {Burger}, {Busca},
  {Campbell}, {Cargile}, {Carithers}, {Carlberg}, {Carr}, {Chang}, {Chen},
  {Chiappini}, {Comparat}, {Connolly}, {Cortes}, {Croft}, {Cunha}, {da Costa},
  {Davenport}, {Dawson}, {De Lee}, {Porto de Mello}, {de Simoni}, {Dean},
  {Dhital}, {Ealet}, {Ebelke}, {Edmondson}, {Eiting}, {Escoffier}, {Esposito},
  {Evans}, {Fan}, {Femen{\'\i}a Castell{\'a}}, {Dutra Ferreira}, {Fitzgerald},
  {Fleming}, {Font-Ribera}, {Ford}, {Frinchaboy}, {Garc{\'\i}a P{\'e}rez},
  {Gaudi}, {Ge}, {Ghezzi}, {Gillespie}, {Gilmore}, {Girardi}, {Gott}, {Gould},
  {Grebel}, {Gunn}, {Hamilton}, {Harding}, {Harris}, {Hawley}, {Hearty},
  {Hennawi}, {Gonz{\'a}lez Hern{\'a}ndez}, {Ho}, {Hogg}, {Holtzman},
  {Honscheid}, {Inada}, {Ivans}, {Jiang}, {Jiang}, {Johnson}, {Jordan},
  {Jordan}, {Kauffmann}, {Kazin}, {Kirkby}, {Klaene}, {Knapp}, {Kneib},
  {Kochanek}, {Koesterke}, {Kollmeier}, {Kron}, {Lampeitl}, {Lang}, {Lawler},
  {Le Goff}, {Lee}, {Lee}, {Leisenring}, {Lin}, {Liu}, {Long}, {Loomis},
  {Lucatello}, {Lundgren}, {Lupton}, {Ma}, {Ma}, {MacDonald}, {Mack},
  {Mahadevan}, {Maia}, {Majewski}, {Makler}, {Malanushenko}, {Malanushenko},
  {Mandelbaum}, {Maraston}, {Margala}, {Maseman}, {Masters}, {McBride},
  {McDonald}, {McGreer}, {McMahon}, {Mena Requejo}, {M{\'e}nard},
  {Miralda-Escud{\'e}}, {Morrison}, {Mullally}, {Muna}, {Murayama}, {Myers},
  {Naugle}, {Neto}, {Nguyen}, {Nichol}, {Nidever}, {O'Connell}, {Ogando},
  {Olmstead}, {Oravetz}, {Padmanabhan}, {Paegert}, {Palanque-Delabrouille},
  {Pan}, {Pandey}, {Parejko}, {P{\^a}ris}, {Pellegrini}, {Pepper}, {Percival},
  {Petitjean}, {Pfaffenberger}, {Pforr}, {Phleps}, {Pichon}, {Pieri}, {Prada},
  {Price-Whelan}, {Raddick}, {Ramos}, {Reid}, {Reyle}, {Rich}, {Richards},
  {Rieke}, {Rieke}, {Rix}, {Robin}, {Rocha-Pinto}, {Rockosi}, {Roe},
  {Rollinde}, {Ross}, {Ross}, {Rossetto}, {S{\'a}nchez}, {Santiago}, {Sayres},
  {Schiavon}, {Schlegel}, {Schlesinger}, {Schmidt}, {Schneider}, {Sellgren},
  {Shelden}, {Sheldon}, {Shetrone}, {Shu}, {Silverman}, {Simmerer}, {Simmons},
  {Sivarani}, {Skrutskie}, {Slosar}, {Smee}, {Smith}, {Snedden}, {Stassun},
  {Steele}, {Steinmetz}, {Stockett}, {Stollberg}, {Strauss}, {Szalay},
  {Tanaka}, {Thakar}, {Thomas}, {Tinker}, {Tofflemire}, {Tojeiro}, {Tremonti},
  {Vargas Maga{\~n}a}, {Verde}, {Vogt}, {Wake}, {Wan}, {Wang}, {Weaver},
  {White}, {White}, {Wilson}, {Wisniewski}, {Wood-Vasey}, {Yanny}, {Yasuda},
  {Y{\`e}che}, {York}, {Young}, {Zasowski}, {Zehavi}, \&
  {Zhao}}]{Eisenstein+11}
{Eisenstein}, D.~J., {Weinberg}, D.~H., {Agol}, E., {et~al.} 2011, \aj, 142,
  72, \dodoi{10.1088/0004-6256/142/3/72}

\bibitem[{{Fardal} {et~al.}(2001){Fardal}, {Katz}, {Gardner}, {Hernquist},
  {Weinberg}, \& {Dav{\'e}}}]{Fardal+01}
{Fardal}, M.~A., {Katz}, N., {Gardner}, J.~P., {et~al.} 2001, \apj, 562, 605,
  \dodoi{10.1086/323519}

\bibitem[{{Faucher-Gigu{\`e}re} {et~al.}(2010){Faucher-Gigu{\`e}re},
  {Kere{\v{s}}}, {Dijkstra}, {Hernquist}, \& {Zaldarriaga}}]{Faucher-Gigure+10}
{Faucher-Gigu{\`e}re}, C.-A., {Kere{\v{s}}}, D., {Dijkstra}, M., {Hernquist},
  L., \& {Zaldarriaga}, M. 2010, \apj, 725, 633,
  \dodoi{10.1088/0004-637X/725/1/633}

\bibitem[{{Feldmeier} {et~al.}(2013){Feldmeier}, {Hagen}, {Ciardullo},
  {Gronwall}, {Gawiser}, {Guaita}, {Hagen}, {Bond}, {Acquaviva}, {Blanc},
  {Orsi}, \& {Kurczynski}}]{Feldmeier+13}
{Feldmeier}, J.~J., {Hagen}, A., {Ciardullo}, R., {et~al.} 2013, \apj, 776, 75,
  \dodoi{10.1088/0004-637X/776/2/75}

\bibitem[{Fisher(1970)}]{Fisher70}
Fisher, Ronald~Aylmer, S. 1970, Statistical methods for research workers, 14th
  edn. (Edinburgh : Oliver and Boyd)

\bibitem[{{Fonseca} {et~al.}(2017){Fonseca}, {Silva}, {Santos}, \&
  {Cooray}}]{Fonseca+17}
{Fonseca}, J., {Silva}, M.~B., {Santos}, M.~G., \& {Cooray}, A. 2017, \mnras,
  464, 1948, \dodoi{10.1093/mnras/stw2470}

\bibitem[{{Fujimoto} {et~al.}(2019){Fujimoto}, {Ouchi}, {Ferrara},
  {Pallottini}, {Ivison}, {Behrens}, {Gallerani}, {Arata}, {Yajima}, \&
  {Nagamine}}]{Fujimoto+19}
{Fujimoto}, S., {Ouchi}, M., {Ferrara}, A., {et~al.} 2019, \apj, 887, 107,
  \dodoi{10.3847/1538-4357/ab480f}

\bibitem[{{Furlanetto} {et~al.}(2005){Furlanetto}, {Schaye}, {Springel}, \&
  {Hernquist}}]{Furlanetto+05}
{Furlanetto}, S.~R., {Schaye}, J., {Springel}, V., \& {Hernquist}, L. 2005,
  \apj, 622, 7, \dodoi{10.1086/426808}

\bibitem[{{Gallego} {et~al.}(2018){Gallego}, {Cantalupo}, {Lilly}, {Marino},
  {Pezzulli}, {Schaye}, {Wisotzki}, {Bacon}, {Inami}, {Akhlaghi}, {Tacchella},
  {Richard}, {Bouche}, {Steinmetz}, \& {Carollo}}]{Gallego+18}
{Gallego}, S.~G., {Cantalupo}, S., {Lilly}, S., {et~al.} 2018, \mnras, 475,
  3854, \dodoi{10.1093/mnras/sty037}

\bibitem[{{Garel} {et~al.}(2021){Garel}, {Blaizot}, {Rosdahl}, {Michel-Dansac},
  {Haehnelt}, {Katz}, {Kimm}, \& {Verhamme}}]{Garel+21}
{Garel}, T., {Blaizot}, J., {Rosdahl}, J., {et~al.} 2021, \mnras, 504, 1902,
  \dodoi{10.1093/mnras/stab990}

\bibitem[{{Goerdt} {et~al.}(2010){Goerdt}, {Dekel}, {Sternberg}, {Ceverino},
  {Teyssier}, \& {Primack}}]{Goerdt+10}
{Goerdt}, T., {Dekel}, A., {Sternberg}, A., {et~al.} 2010, \mnras, 407, 613,
  \dodoi{10.1111/j.1365-2966.2010.16941.x}

\bibitem[{{Gong} {et~al.}(2011){Gong}, {Cooray}, {Silva}, {Santos}, \&
  {Lubin}}]{Gong+11}
{Gong}, Y., {Cooray}, A., {Silva}, M.~B., {Santos}, M.~G., \& {Lubin}, P. 2011,
  \apjl, 728, L46, \dodoi{10.1088/2041-8205/728/2/L46}

\bibitem[{{Goto} {et~al.}(2009){Goto}, {Utsumi}, {Furusawa}, {Miyazaki}, \&
  {Komiyama}}]{Goto+09}
{Goto}, T., {Utsumi}, Y., {Furusawa}, H., {Miyazaki}, S., \& {Komiyama}, Y.
  2009, \mnras, 400, 843, \dodoi{10.1111/j.1365-2966.2009.15486.x}

\bibitem[{{Haiman} {et~al.}(2000){Haiman}, {Spaans}, \& {Quataert}}]{Haiman+00}
{Haiman}, Z., {Spaans}, M., \& {Quataert}, E. 2000, \apjl, 537, L5,
  \dodoi{10.1086/312754}

\bibitem[{{Harikane} {et~al.}(2016){Harikane}, {Ouchi}, {Ono}, {More}, {Saito},
  {Lin}, {Coupon}, {Shimasaku}, {Shibuya}, {Price}, {Lin}, {Hsieh}, {Ishigaki},
  {Komiyama}, {Silverman}, {Takata}, {Tamazawa}, \& {Toshikawa}}]{Harikane+16}
{Harikane}, Y., {Ouchi}, M., {Ono}, Y., {et~al.} 2016, \apj, 821, 123,
  \dodoi{10.3847/0004-637X/821/2/123}

\bibitem[{{Harikane} {et~al.}(2018){Harikane}, {Ouchi}, {Shibuya}, {Kojima},
  {Zhang}, {Itoh}, {Ono}, {Higuchi}, {Inoue}, {Chevallard}, {Capak}, {Nagao},
  {Onodera}, {Faisst}, {Martin}, {Rauch}, {Bruzual}, {Charlot}, {Davidzon},
  {Fujimoto}, {Hilmi}, {Ilbert}, {Lee}, {Matsuoka}, {Silverman}, \&
  {Toft}}]{Harikane+18b}
{Harikane}, Y., {Ouchi}, M., {Shibuya}, T., {et~al.} 2018, \apj, 859, 84,
  \dodoi{10.3847/1538-4357/aabd80}

\bibitem[{{Harikane} {et~al.}(2019){Harikane}, {Ouchi}, {Ono}, {Fujimoto},
  {Donevski}, {Shibuya}, {Faisst}, {Goto}, {Hatsukade}, {Kashikawa}, {Kohno},
  {Hashimoto}, {Higuchi}, {Inoue}, {Lin}, {Martin}, {Overzier}, {Smail},
  {Toshikawa}, {Umehata}, {Ao}, {Chapman}, {Clements}, {Im}, {Jing},
  {Kawaguchi}, {Lee}, {Lee}, {Lin}, {Matsuoka}, {Marinello}, {Nagao},
  {Onodera}, {Toft}, \& {Wang}}]{Harikane+19}
{Harikane}, Y., {Ouchi}, M., {Ono}, Y., {et~al.} 2019, \apj, 883, 142,
  \dodoi{10.3847/1538-4357/ab2cd5}

\bibitem[{{Hayashi} {et~al.}(2020){Hayashi}, {Shimakawa}, {Tanaka}, {Onodera},
  {Koyama}, {Inoue}, {Komiyama}, {Lee}, {Lin}, \& {Yabe}}]{Hayashi+20}
{Hayashi}, M., {Shimakawa}, R., {Tanaka}, M., {et~al.} 2020, \pasj, 72, 86,
  \dodoi{10.1093/pasj/psaa076}

\bibitem[{{Hayashino} {et~al.}(2004){Hayashino}, {Matsuda}, {Tamura},
  {Yamauchi}, {Yamada}, {Ajiki}, {Fujita}, {Murayama}, {Nagao}, {Ohta},
  {Okamura}, {Ouchi}, {Shimasaku}, {Shioya}, \& {Taniguchi}}]{Hayashino+04}
{Hayashino}, T., {Matsuda}, Y., {Tamura}, H., {et~al.} 2004, \aj, 128, 2073,
  \dodoi{10.1086/424935}

\bibitem[{{Hayes} {et~al.}(2011){Hayes}, {Schaerer}, {{\"O}stlin}, {Mas-Hesse},
  {Atek}, \& {Kunth}}]{Hayes+11}
{Hayes}, M., {Schaerer}, D., {{\"O}stlin}, G., {et~al.} 2011, \apj, 730, 8,
  \dodoi{10.1088/0004-637X/730/1/8}

\bibitem[{{Hayes} {et~al.}(2013){Hayes}, {{\"O}stlin}, {Schaerer}, {Verhamme},
  {Mas-Hesse}, {Adamo}, {Atek}, {Cannon}, {Duval}, {Guaita}, {Herenz}, {Kunth},
  {Laursen}, {Melinder}, {Orlitov{\'a}}, {Ot{\'\i}-Floranes}, \&
  {Sandberg}}]{Hayes+13}
{Hayes}, M., {{\"O}stlin}, G., {Schaerer}, D., {et~al.} 2013, \apjl, 765, L27,
  \dodoi{10.1088/2041-8205/765/2/L27}

\bibitem[{{Hayes} {et~al.}(2014){Hayes}, {{\"O}stlin}, {Duval}, {Sandberg},
  {Guaita}, {Melinder}, {Adamo}, {Schaerer}, {Verhamme}, {Orlitov{\'a}},
  {Mas-Hesse}, {Cannon}, {Atek}, {Kunth}, {Laursen}, {Ot{\'\i}-Floranes},
  {Pardy}, {Rivera-Thorsen}, \& {Herenz}}]{Hayes+14}
{Hayes}, M., {{\"O}stlin}, G., {Duval}, F., {et~al.} 2014, \apj, 782, 6,
  \dodoi{10.1088/0004-637X/782/1/6}

\bibitem[{{Higuchi} {et~al.}(2019){Higuchi}, {Ouchi}, {Ono}, {Shibuya},
  {Toshikawa}, {Harikane}, {Kojima}, {Chiang}, {Egami}, {Kashikawa},
  {Overzier}, {Konno}, {Inoue}, {Hasegawa}, {Fujimoto}, {Goto}, {Ishikawa},
  {Ito}, {Komiyama}, \& {Tanaka}}]{Higuchi+18}
{Higuchi}, R., {Ouchi}, M., {Ono}, Y., {et~al.} 2019, \apj, 879, 28,
  \dodoi{10.3847/1538-4357/ab2192}

\bibitem[{{Inoue} {et~al.}(2014){Inoue}, {Shimizu}, {Iwata}, \&
  {Tanaka}}]{Inoue+14}
{Inoue}, A.~K., {Shimizu}, I., {Iwata}, I., \& {Tanaka}, M. 2014, \mnras, 442,
  1805, \dodoi{10.1093/mnras/stu936}

\bibitem[{{Inoue} {et~al.}(2018){Inoue}, {Hasegawa}, {Ishiyama}, {Yajima},
  {Shimizu}, {Umemura}, {Konno}, {Harikane}, {Shibuya}, {Ouchi}, {Shimasaku},
  {Ono}, {Kusakabe}, {Higuchi}, \& {Lee}}]{Inoue+18}
{Inoue}, A.~K., {Hasegawa}, K., {Ishiyama}, T., {et~al.} 2018, \pasj, 70, 55,
  \dodoi{10.1093/pasj/psy048}

\bibitem[{{Inoue} {et~al.}(2020){Inoue}, {Yamanaka}, {Ouchi}, {Iwata},
  {Shimasaku}, {Taniguchi}, {Nagao}, {Kashikawa}, {Ono}, {Mawatari}, {Shibuya},
  {Hayashi}, {Ikeda}, {Zhang}, {Liang}, {Lee}, {Hilmi}, {Kikuta}, {Kusakabe},
  {Furusawa}, {Hayashino}, {Kajisawa}, {Matsuda}, {Nakajima}, {Momose},
  {Harikane}, {Saito}, {Kodama}, {Kikuchihara}, {Iye}, \& {Goto}}]{Inoue+20}
{Inoue}, A.~K., {Yamanaka}, S., {Ouchi}, M., {et~al.} 2020, \pasj, 72, 101,
  \dodoi{10.1093/pasj/psaa100}

\bibitem[{{Itoh} {et~al.}(2018){Itoh}, {Ouchi}, {Zhang}, {Inoue}, {Mawatari},
  {Shibuya}, {Harikane}, {Ono}, {Kusakabe}, {Shimasaku}, {Fujimoto}, {Iwata},
  {Kajisawa}, {Kashikawa}, {Kawanomoto}, {Komiyama}, {Lee}, {Nagao}, \&
  {Taniguchi}}]{Itoh+18}
{Itoh}, R., {Ouchi}, M., {Zhang}, H., {et~al.} 2018, \apj, 867, 46,
  \dodoi{10.3847/1538-4357/aadfe4}

\bibitem[{{Jeeson-Daniel} {et~al.}(2012){Jeeson-Daniel}, {Ciardi}, {Maio},
  {Pierleoni}, {Dijkstra}, \& {Maselli}}]{Jeeson-Daniel+12}
{Jeeson-Daniel}, A., {Ciardi}, B., {Maio}, U., {et~al.} 2012, \mnras, 424,
  2193, \dodoi{10.1111/j.1365-2966.2012.21378.x}

\bibitem[{{Jiang} {et~al.}(2013){Jiang}, {Egami}, {Fan}, {Windhorst}, {Cohen},
  {Dav{\'e}}, {Finlator}, {Kashikawa}, {Mechtley}, {Ouchi}, \&
  {Shimasaku}}]{Jiang+13}
{Jiang}, L., {Egami}, E., {Fan}, X., {et~al.} 2013, \apj, 773, 153,
  \dodoi{10.1088/0004-637X/773/2/153}

\bibitem[{{Kakiichi} \& {Dijkstra}(2018)}]{Kakiichi-Dijkstra18}
{Kakiichi}, K., \& {Dijkstra}, M. 2018, \mnras, 480, 5140,
  \dodoi{10.1093/mnras/sty2214}

\bibitem[{{Kakuma} {et~al.}(2021){Kakuma}, {Ouchi}, {Harikane}, {Inoue},
  {Komiyama}, {Kusakabe}, {Liu}, {Matsuda}, {Matsuoka}, {Mawatari}, {Momose},
  {Ono}, {Shibuya}, \& {Taniguchi}}]{Kakuma+21}
{Kakuma}, R., {Ouchi}, M., {Harikane}, Y., {et~al.} 2021, arXiv e-prints,
  arXiv:1906.00173.
\newblock \doarXiv{1906.00173}

\bibitem[{{Kikuta} {et~al.}(2019){Kikuta}, {Matsuda}, {Cen}, {Steidel}, {Yagi},
  {Hayashino}, {Imanishi}, {Komiyama}, {Momose}, \& {Saito}}]{Kikuta+19}
{Kikuta}, S., {Matsuda}, Y., {Cen}, R., {et~al.} 2019, \pasj, 71, L2,
  \dodoi{10.1093/pasj/psz055}

\bibitem[{{Kollmeier} {et~al.}(2010){Kollmeier}, {Zheng}, {Dav{\'e}}, {Gould},
  {Katz}, {Miralda-Escud{\'e}}, \& {Weinberg}}]{Kollmeier+10}
{Kollmeier}, J.~A., {Zheng}, Z., {Dav{\'e}}, R., {et~al.} 2010, \apj, 708,
  1048, \dodoi{10.1088/0004-637X/708/2/1048}

\bibitem[{{Konno} {et~al.}(2016){Konno}, {Ouchi}, {Nakajima}, {Duval},
  {Kusakabe}, {Ono}, \& {Shimasaku}}]{Konno+16}
{Konno}, A., {Ouchi}, M., {Nakajima}, K., {et~al.} 2016, \apj, 823, 20,
  \dodoi{10.3847/0004-637X/823/1/20}

\bibitem[{{Konno} {et~al.}(2018){Konno}, {Ouchi}, {Shibuya}, {Ono},
  {Shimasaku}, {Taniguchi}, {Nagao}, {Kobayashi}, {Kajisawa}, {Kashikawa},
  {Inoue}, {Oguri}, {Furusawa}, {Goto}, {Harikane}, {Higuchi}, {Komiyama},
  {Kusakabe}, {Miyazaki}, {Nakajima}, \& {Wang}}]{Konno+18}
{Konno}, A., {Ouchi}, M., {Shibuya}, T., {et~al.} 2018, \pasj, 70, S16,
  \dodoi{10.1093/pasj/psx131}

\bibitem[{{Kovetz} {et~al.}(2017){Kovetz}, {Viero}, {Lidz}, {Newburgh},
  {Rahman}, {Switzer}, {Kamionkowski}, {Aguirre}, {Alvarez}, {Bock}, {Bond},
  {Bower}, {Bradford}, {Breysse}, {Bull}, {Chang}, {Cheng}, {Chung}, {Cleary},
  {Corray}, {Crites}, {Croft}, {Dor{\'e}}, {Eastwood}, {Ferrara}, {Fonseca},
  {Jacobs}, {Keating}, {Lagache}, {Lakhlani}, {Liu}, {Moodley}, {Murray},
  {P{\'e}nin}, {Popping}, {Pullen}, {Reichers}, {Saito}, {Saliwanchik},
  {Santos}, {Somerville}, {Stacey}, {Stein}, {Villaescusa-Navarro}, {Visbal},
  {Weltman}, {Wolz}, \& {Zemcov}}]{Kovetz+17}
{Kovetz}, E.~D., {Viero}, M.~P., {Lidz}, A., {et~al.} 2017, arXiv e-prints,
  arXiv:1709.09066.
\newblock \doarXiv{1709.09066}

\bibitem[{{Kusakabe} {et~al.}(2018){Kusakabe}, {Shimasaku}, {Ouchi},
  {Nakajima}, {Goto}, {Hashimoto}, {Konno}, {Harikane}, {Silverman}, \&
  {Capak}}]{Kusakabe+18}
{Kusakabe}, H., {Shimasaku}, K., {Ouchi}, M., {et~al.} 2018, \pasj, 70, 4,
  \dodoi{10.1093/pasj/psx148}

\bibitem[{{Kusakabe} {et~al.}(2019){Kusakabe}, {Shimasaku}, {Momose}, {Ouchi},
  {Nakajima}, {Hashimoto}, {Harikane}, {Silverman}, \& {Capak}}]{Kusakabe+19}
{Kusakabe}, H., {Shimasaku}, K., {Momose}, R., {et~al.} 2019, \pasj, 71, 55,
  \dodoi{10.1093/pasj/psz029}

\bibitem[{{Lake} {et~al.}(2015){Lake}, {Zheng}, {Cen}, {Sadoun}, {Momose}, \&
  {Ouchi}}]{Lake+15}
{Lake}, E., {Zheng}, Z., {Cen}, R., {et~al.} 2015, \apj, 806, 46,
  \dodoi{10.1088/0004-637X/806/1/46}

\bibitem[{{Laursen} \& {Sommer-Larsen}(2007)}]{Laursen-Sommer-Larsen07}
{Laursen}, P., \& {Sommer-Larsen}, J. 2007, \apjl, 657, L69,
  \dodoi{10.1086/513191}

\bibitem[{{Laursen} {et~al.}(2011){Laursen}, {Sommer-Larsen}, \&
  {Razoumov}}]{Laursen+11}
{Laursen}, P., {Sommer-Larsen}, J., \& {Razoumov}, A.~O. 2011, \apj, 728, 52,
  \dodoi{10.1088/0004-637X/728/1/52}

\bibitem[{{Leclercq} {et~al.}(2017){Leclercq}, {Bacon}, {Wisotzki}, {Mitchell},
  {Garel}, {Verhamme}, {Blaizot}, {Hashimoto}, {Herenz}, {Conseil},
  {Cantalupo}, {Inami}, {Contini}, {Richard}, {Maseda}, {Schaye}, {Marino},
  {Akhlaghi}, {Brinchmann}, \& {Carollo}}]{Leclercq+17}
{Leclercq}, F., {Bacon}, R., {Wisotzki}, L., {et~al.} 2017, \aap, 608, A8,
  \dodoi{10.1051/0004-6361/201731480}

\bibitem[{{Leclercq} {et~al.}(2020){Leclercq}, {Bacon}, {Verhamme}, {Garel},
  {Blaizot}, {Brinchmann}, {Cantalupo}, {Claeyssens}, {Conseil}, {Contini},
  {Hashimoto}, {Herenz}, {Kusakabe}, {Marino}, {Maseda}, {Matthee}, {Mitchell},
  {Pezzulli}, {Richard}, {Schmidt}, \& {Wisotzki}}]{Leclercq+20}
{Leclercq}, F., {Bacon}, R., {Verhamme}, A., {et~al.} 2020, \aap, 635, A82,
  \dodoi{10.1051/0004-6361/201937339}

\bibitem[{{Li} {et~al.}(2016){Li}, {Wechsler}, {Devaraj}, \& {Church}}]{Li+16}
{Li}, T.~Y., {Wechsler}, R.~H., {Devaraj}, K., \& {Church}, S.~E. 2016, \apj,
  817, 169, \dodoi{10.3847/0004-637X/817/2/169}

\bibitem[{{Martin} {et~al.}(2014){Martin}, {Chang}, {Matuszewski}, {Morrissey},
  {Rahman}, {Moore}, {Steidel}, \& {Matsuda}}]{Martin+14}
{Martin}, D.~C., {Chang}, D., {Matuszewski}, M., {et~al.} 2014, \apj, 786, 107,
  \dodoi{10.1088/0004-637X/786/2/107}

\bibitem[{{Mas-Ribas} \& {Dijkstra}(2016)}]{Mas-Ribas-Dijkstra16}
{Mas-Ribas}, L., \& {Dijkstra}, M. 2016, \apj, 822, 84,
  \dodoi{10.3847/0004-637X/822/2/84}

\bibitem[{{Mas-Ribas} {et~al.}(2017{\natexlab{a}}){Mas-Ribas}, {Dijkstra},
  {Hennawi}, {Trenti}, {Momose}, \& {Ouchi}}]{Mas-Ribas+17a}
{Mas-Ribas}, L., {Dijkstra}, M., {Hennawi}, J.~F., {et~al.} 2017{\natexlab{a}},
  \apj, 841, 19, \dodoi{10.3847/1538-4357/aa704e}

\bibitem[{{Mas-Ribas} {et~al.}(2017{\natexlab{b}}){Mas-Ribas}, {Hennawi},
  {Dijkstra}, {Davies}, {Stern}, \& {Rix}}]{Mas-Ribas+17c}
{Mas-Ribas}, L., {Hennawi}, J.~F., {Dijkstra}, M., {et~al.} 2017{\natexlab{b}},
  \apj, 846, 11, \dodoi{10.3847/1538-4357/aa8328}

\bibitem[{{Matsuda} {et~al.}(2012){Matsuda}, {Yamada}, {Hayashino}, {Yamauchi},
  {Nakamura}, {Morimoto}, {Ouchi}, {Ono}, {Umemura}, \& {Mori}}]{Matsuda+12}
{Matsuda}, Y., {Yamada}, T., {Hayashino}, T., {et~al.} 2012, \mnras, 425, 878,
  \dodoi{10.1111/j.1365-2966.2012.21143.x}

\bibitem[{{Mitchell} {et~al.}(2021){Mitchell}, {Blaizot}, {Cadiou}, {Dubois},
  {Garel}, \& {Rosdahl}}]{Mitchell+21}
{Mitchell}, P.~D., {Blaizot}, J., {Cadiou}, C., {et~al.} 2021, \mnras, 501,
  5757, \dodoi{10.1093/mnras/stab035}

\bibitem[{{Momose} {et~al.}(2021{\natexlab{a}}){Momose}, {Shimasaku},
  {Nagamine}, {Shimizu}, {Kashikawa}, {Ando}, \& {Kusakabe}}]{Momose+21b}
{Momose}, R., {Shimasaku}, K., {Nagamine}, K., {et~al.} 2021{\natexlab{a}},
  \apjl, 912, L24, \dodoi{10.3847/2041-8213/abf04c}

\bibitem[{{Momose} {et~al.}(2021{\natexlab{b}}){Momose}, {Shimizu}, {Nagamine},
  {Shimasaku}, {Kashikawa}, \& {Kusakabe}}]{Momose+21a}
{Momose}, R., {Shimizu}, I., {Nagamine}, K., {et~al.} 2021{\natexlab{b}}, \apj,
  911, 98, \dodoi{10.3847/1538-4357/abe1b9}

\bibitem[{{Momose} {et~al.}(2014){Momose}, {Ouchi}, {Nakajima}, {Ono},
  {Shibuya}, {Shimasaku}, {Yuma}, {Mori}, \& {Umemura}}]{Momose+14}
{Momose}, R., {Ouchi}, M., {Nakajima}, K., {et~al.} 2014, \mnras, 442, 110,
  \dodoi{10.1093/mnras/stu825}

\bibitem[{{Momose} {et~al.}(2016){Momose}, {Ouchi}, {Nakajima}, {Ono},
  {Shibuya}, {Shimasaku}, {Yuma}, {Mori}, \& {Umemura}}]{Momose+16}
---. 2016, \mnras, 457, 2318, \dodoi{10.1093/mnras/stw021}

\bibitem[{{Navarro} {et~al.}(1997){Navarro}, {Frenk}, \& {White}}]{Navarro+97}
{Navarro}, J.~F., {Frenk}, C.~S., \& {White}, S. D.~M. 1997, \apj, 490, 493,
  \dodoi{10.1086/304888}

\bibitem[{{Oke} \& {Gunn}(1983)}]{Oke-Gunn83}
{Oke}, J.~B., \& {Gunn}, J.~E. 1983, \apj, 266, 713, \dodoi{10.1086/160817}

\bibitem[{{Ono} {et~al.}(2021){Ono}, {Itoh}, {Shibuya}, {Ouchi}, {Harikane},
  {Yamanaka}, {Inoue}, {Amagasa}, {Miura}, {Okura}, {Shimasaku}, {Iwata},
  {Taniguchi}, {Fujimoto}, {Iye}, {Jaelani}, {Kashikawa}, {Kikuchihara},
  {Kikuta}, {Kobayashi}, {Kusakabe}, {Lee}, {Liang}, {Matsuoka}, {Momose},
  {Nagao}, {Nakajima}, \& {Tadaki}}]{Ono+21}
{Ono}, Y., {Itoh}, R., {Shibuya}, T., {et~al.} 2021, \apj, 911, 78,
  \dodoi{10.3847/1538-4357/abea15}

\bibitem[{{{\"O}stlin} {et~al.}(2009){{\"O}stlin}, {Hayes}, {Kunth},
  {Mas-Hesse}, {Leitherer}, {Petrosian}, \& {Atek}}]{Ostlin+09}
{{\"O}stlin}, G., {Hayes}, M., {Kunth}, D., {et~al.} 2009, \aj, 138, 923,
  \dodoi{10.1088/0004-6256/138/3/923}

\bibitem[{{Ouchi} {et~al.}(2020){Ouchi}, {Ono}, \& {Shibuya}}]{Ouchi+20}
{Ouchi}, M., {Ono}, Y., \& {Shibuya}, T. 2020, \araa, 58, 617,
  \dodoi{10.1146/annurev-astro-032620-021859}

\bibitem[{{Ouchi} {et~al.}(2010){Ouchi}, {Shimasaku}, {Furusawa}, {Saito},
  {Yoshida}, {Akiyama}, {Ono}, {Yamada}, {Ota}, {Kashikawa}, {Iye}, {Kodama},
  {Okamura}, {Simpson}, \& {Yoshida}}]{Ouchi+10}
{Ouchi}, M., {Shimasaku}, K., {Furusawa}, H., {et~al.} 2010, \apj, 723, 869,
  \dodoi{10.1088/0004-637X/723/1/869}

\bibitem[{{Ouchi} {et~al.}(2018){Ouchi}, {Harikane}, {Shibuya}, {Shimasaku},
  {Taniguchi}, {Konno}, {Kobayashi}, {Kajisawa}, {Nagao}, {Ono}, {Inoue},
  {Umemura}, {Mori}, {Hasegawa}, {Higuchi}, {Komiyama}, {Matsuda}, {Nakajima},
  {Saito}, \& {Wang}}]{Ouchi+18}
{Ouchi}, M., {Harikane}, Y., {Shibuya}, T., {et~al.} 2018, \pasj, 70, S13,
  \dodoi{10.1093/pasj/psx074}

\bibitem[{{Patr{\'\i}cio} {et~al.}(2016){Patr{\'\i}cio}, {Richard}, {Verhamme},
  {Wisotzki}, {Brinchmann}, {Turner}, {Christensen}, {Weilbacher}, {Blaizot},
  {Bacon}, {Contini}, {Lagattuta}, {Cantalupo}, {Cl{\'e}ment}, \&
  {Soucail}}]{Patricio+16}
{Patr{\'\i}cio}, V., {Richard}, J., {Verhamme}, A., {et~al.} 2016, \mnras, 456,
  4191, \dodoi{10.1093/mnras/stv2859}

\bibitem[{{P{\'e}roux} \& {Howk}(2020)}]{Peroux-Howk20}
{P{\'e}roux}, C., \& {Howk}, J.~C. 2020, \araa, 58, 363,
  \dodoi{10.1146/annurev-astro-021820-120014}

\bibitem[{{Pullen} {et~al.}(2014){Pullen}, {Dor{\'e}}, \& {Bock}}]{Pullen+14}
{Pullen}, A.~R., {Dor{\'e}}, O., \& {Bock}, J. 2014, \apj, 786, 111,
  \dodoi{10.1088/0004-637X/786/2/111}

\bibitem[{{Rauch} {et~al.}(2008){Rauch}, {Haehnelt}, {Bunker}, {Becker},
  {Marleau}, {Graham}, {Cristiani}, {Jarvis}, {Lacey}, {Morris}, {Peroux},
  {R{\"o}ttgering}, \& {Theuns}}]{Rauch+08}
{Rauch}, M., {Haehnelt}, M., {Bunker}, A., {et~al.} 2008, \apj, 681, 856,
  \dodoi{10.1086/525846}

\bibitem[{{Rosdahl} \& {Blaizot}(2012)}]{Rosdahl-Blaizot12}
{Rosdahl}, J., \& {Blaizot}, J. 2012, \mnras, 423, 344,
  \dodoi{10.1111/j.1365-2966.2012.20883.x}

\bibitem[{{Scoville} {et~al.}(2007){Scoville}, {Aussel}, {Brusa}, {Capak},
  {Carollo}, {Elvis}, {Giavalisco}, {Guzzo}, {Hasinger}, {Impey}, {Kneib},
  {LeFevre}, {Lilly}, {Mobasher}, {Renzini}, {Rich}, {Sanders}, {Schinnerer},
  {Schminovich}, {Shopbell}, {Taniguchi}, \& {Tyson}}]{Scoville+07}
{Scoville}, N., {Aussel}, H., {Brusa}, M., {et~al.} 2007, \apjs, 172, 1,
  \dodoi{10.1086/516585}

\bibitem[{{Sekiguchi} {et~al.}(2005){Sekiguchi}, {Akiyama}, {Furusawa},
  {Simpson}, {Takata}, {Ueda}, {Watson}, \& {Sxds Team}}]{Sekiguchi+05}
{Sekiguchi}, K., {Akiyama}, M., {Furusawa}, H., {et~al.} 2005, in
  Multiwavelength Mapping of Galaxy Formation and Evolution, ed. A.~{Renzini}
  \& R.~{Bender}, 82, \dodoi{10.1007/10995020\_12}

\bibitem[{{Shibuya} {et~al.}(2018{\natexlab{a}}){Shibuya}, {Ouchi}, {Konno},
  {Higuchi}, {Harikane}, {Ono}, {Shimasaku}, {Taniguchi}, {Kobayashi},
  {Kajisawa}, {Nagao}, {Furusawa}, {Goto}, {Kashikawa}, {Komiyama}, {Kusakabe},
  {Lee}, {Momose}, {Nakajima}, {Tanaka}, {Wang}, \& {Yuma}}]{Shibuya+18a}
{Shibuya}, T., {Ouchi}, M., {Konno}, A., {et~al.} 2018{\natexlab{a}}, \pasj,
  70, S14, \dodoi{10.1093/pasj/psx122}

\bibitem[{{Shibuya} {et~al.}(2018{\natexlab{b}}){Shibuya}, {Ouchi}, {Harikane},
  {Rauch}, {Ono}, {Mukae}, {Higuchi}, {Kojima}, {Yuma}, {Lee}, {Furusawa},
  {Konno}, {Martin}, {Shimasaku}, {Taniguchi}, {Kobayashi}, {Kajisawa},
  {Nagao}, {Goto}, {Kashikawa}, {Komiyama}, {Kusakabe}, {Momose}, {Nakajima},
  {Tanaka}, \& {Wang}}]{Shibuya+18b}
{Shibuya}, T., {Ouchi}, M., {Harikane}, Y., {et~al.} 2018{\natexlab{b}}, \pasj,
  70, S15, \dodoi{10.1093/pasj/psx107}

\bibitem[{{Silva} {et~al.}(2013){Silva}, {Santos}, {Gong}, {Cooray}, \&
  {Bock}}]{Silva+13}
{Silva}, M.~B., {Santos}, M.~G., {Gong}, Y., {Cooray}, A., \& {Bock}, J. 2013,
  \apj, 763, 132, \dodoi{10.1088/0004-637X/763/2/132}

\bibitem[{{Smit} {et~al.}(2017){Smit}, {Swinbank}, {Massey}, {Richard},
  {Smail}, \& {Kneib}}]{Smit+17}
{Smit}, R., {Swinbank}, A.~M., {Massey}, R., {et~al.} 2017, \mnras, 467, 3306,
  \dodoi{10.1093/mnras/stx245}

\bibitem[{{Smith} {et~al.}(2019){Smith}, {Ma}, {Bromm}, {Finkelstein},
  {Hopkins}, {Faucher-Gigu{\`e}re}, \& {Kere{\v{s}}}}]{Smith+19}
{Smith}, A., {Ma}, X., {Bromm}, V., {et~al.} 2019, \mnras, 484, 39,
  \dodoi{10.1093/mnras/sty3483}

\bibitem[{{Smith} {et~al.}(2018){Smith}, {Tsang}, {Bromm}, \&
  {Milosavljevi{\'c}}}]{Smith+18}
{Smith}, A., {Tsang}, B. T.~H., {Bromm}, V., \& {Milosavljevi{\'c}}, M. 2018,
  \mnras, 479, 2065, \dodoi{10.1093/mnras/sty1509}

\bibitem[{{Steidel} {et~al.}(2011){Steidel}, {Bogosavljevi{\'c}}, {Shapley},
  {Kollmeier}, {Reddy}, {Erb}, \& {Pettini}}]{Steidel+11}
{Steidel}, C.~C., {Bogosavljevi{\'c}}, M., {Shapley}, A.~E., {et~al.} 2011,
  \apj, 736, 160, \dodoi{10.1088/0004-637X/736/2/160}

\bibitem[{{Steidel} {et~al.}(2010){Steidel}, {Erb}, {Shapley}, {Pettini},
  {Reddy}, {Bogosavljevi{\'c}}, {Rudie}, \& {Rakic}}]{Steidel+10}
{Steidel}, C.~C., {Erb}, D.~K., {Shapley}, A.~E., {et~al.} 2010, \apj, 717,
  289, \dodoi{10.1088/0004-637X/717/1/289}

\bibitem[{{Swinbank} {et~al.}(2007){Swinbank}, {Bower}, {Smith}, {Wilman},
  {Smail}, {Ellis}, {Morris}, \& {Kneib}}]{Swinbank+07}
{Swinbank}, A.~M., {Bower}, R.~G., {Smith}, G.~P., {et~al.} 2007, \mnras, 376,
  479, \dodoi{10.1111/j.1365-2966.2007.11454.x}

\bibitem[{{Tumlinson} {et~al.}(2017){Tumlinson}, {Peeples}, \&
  {Werk}}]{Tumlinson+17}
{Tumlinson}, J., {Peeples}, M.~S., \& {Werk}, J.~K. 2017, \araa, 55, 389,
  \dodoi{10.1146/annurev-astro-091916-055240}

\bibitem[{{Verhamme} {et~al.}(2012){Verhamme}, {Dubois}, {Blaizot}, {Garel},
  {Bacon}, {Devriendt}, {Guiderdoni}, \& {Slyz}}]{Verhamme+12}
{Verhamme}, A., {Dubois}, Y., {Blaizot}, J., {et~al.} 2012, \aap, 546, A111,
  \dodoi{10.1051/0004-6361/201218783}

\bibitem[{{Wisotzki} {et~al.}(2016){Wisotzki}, {Bacon}, {Blaizot},
  {Brinchmann}, {Herenz}, {Schaye}, {Bouch{\'e}}, {Cantalupo}, {Contini},
  {Carollo}, {Caruana}, {Courbot}, {Emsellem}, {Kamann}, {Kerutt}, {Leclercq},
  {Lilly}, {Patr{\'\i}cio}, {Sandin}, {Steinmetz}, {Straka}, {Urrutia},
  {Verhamme}, {Weilbacher}, \& {Wendt}}]{Wisotzki+16}
{Wisotzki}, L., {Bacon}, R., {Blaizot}, J., {et~al.} 2016, \aap, 587, A98,
  \dodoi{10.1051/0004-6361/201527384}

\bibitem[{{Wisotzki} {et~al.}(2018){Wisotzki}, {Bacon}, {Brinchmann},
  {Cantalupo}, {Richter}, {Schaye}, {Schmidt}, {Urrutia}, {Weilbacher},
  {Akhlaghi}, {Bouch{\'e}}, {Contini}, {Guiderdoni}, {Herenz}, {Inami},
  {Kerutt}, {Leclercq}, {Marino}, {Maseda}, {Monreal-Ibero}, {Nanayakkara},
  {Richard}, {Saust}, {Steinmetz}, \& {Wendt}}]{Wisotzki+18}
{Wisotzki}, L., {Bacon}, R., {Brinchmann}, J., {et~al.} 2018, \nat, 562, 229,
  \dodoi{10.1038/s41586-018-0564-6}

\bibitem[{{Wu} {et~al.}(2020){Wu}, {Jiang}, \& {Ning}}]{Wu+20}
{Wu}, J., {Jiang}, L., \& {Ning}, Y. 2020, \apj, 891, 105,
  \dodoi{10.3847/1538-4357/ab7333}

\bibitem[{{Xue} {et~al.}(2017){Xue}, {Lee}, {Dey}, {Reddy}, {Hong}, {Prescott},
  {Inami}, {Jannuzi}, \& {Gonzalez}}]{Xue+17}
{Xue}, R., {Lee}, K.-S., {Dey}, A., {et~al.} 2017, \apj, 837, 172,
  \dodoi{10.3847/1538-4357/837/2/172}

\bibitem[{{Zhang} {et~al.}(2020){Zhang}, {Ouchi}, {Itoh}, {Shibuya}, {Ono},
  {Harikane}, {Inoue}, {Rauch}, {Kikuchihara}, {Nakajima}, {Yajima}, {Arata},
  {Abe}, {Iwata}, {Kashikawa}, {Kawanomoto}, {Kikuta}, {Kobayashi}, {Kusakabe},
  {Mawatari}, {Nagao}, {Shimasaku}, \& {Taniguchi}}]{Zhang+20}
{Zhang}, H., {Ouchi}, M., {Itoh}, R., {et~al.} 2020, \apj, 891, 177,
  \dodoi{10.3847/1538-4357/ab7917}

\bibitem[{{Zheng} {et~al.}(2011){Zheng}, {Cen}, {Weinberg}, {Trac}, \&
  {Miralda-Escud{\'e}}}]{Zheng+11}
{Zheng}, Z., {Cen}, R., {Weinberg}, D., {Trac}, H., \& {Miralda-Escud{\'e}}, J.
  2011, \apj, 739, 62, \dodoi{10.1088/0004-637X/739/2/62}

\end{thebibliography}
\end{document}